\documentclass[Journal]{IEEEtran}
\usepackage{amsfonts}
\IEEEoverridecommandlockouts

\ifCLASSINFOpdf
\else
\fi
\usepackage{subfigure}
\usepackage{epstopdf}
\usepackage{multirow}
\usepackage{cite}
\usepackage{stfloats}
\usepackage{color}
\usepackage{epsfig}
\usepackage{graphicx}
\usepackage{psfig}
\usepackage{epsf}
\usepackage{float}
\usepackage{amssymb }
\usepackage[cmex10]{amsmath}
\usepackage{algorithm} %format of the algorithm
\usepackage{algorithmic}
\usepackage{booktabs}
\usepackage{amsmath,bm}
\usepackage{tabularx}
\usepackage{array}
\usepackage{float}
\usepackage{subfigure}
\usepackage{hyperref}
\usepackage{tikz,xcolor}
\begin{document}
\title{Symbol Detection  in Ambient Backscatter Communications Under Residual Time Synchronization Errors}
\author{Yinghui Ye, Ying Li, Xiaoli Chu,~\IEEEmembership{Senior Member,~IEEE}, Gan Zheng,~\IEEEmembership{Fellow,~IEEE}, and Sumei Sun,~\IEEEmembership{Fellow,~IEEE}\vspace{-25pt}
 \thanks{Yinghui Ye and Ying Li   are with the Shaanxi Key Laboratory of
Information Communication Network and Security, Xi'an University of Posts
\& Telecommunications, China. Xiaoli Chu is with the Department of Electronic and Electrical Engineering, The University of Sheffield, Sheffield, U.K. Gan Zheng is with the School of Engineering, University of Warwick, Coventry, U.K.  Sumei Sun is with the Institute of Infocomm Research, Agency for Science, Technology and Research, Singapore.}%(e-mail: 15229972187@163.com, connectyyh@126.com)
%\thanks{ }
%\thanks{}
%\thanks{}
%\thanks{This work was supported  in part by  the National Natural Science Foundation of China under Grant XXXX,  in part by the Young Talent fund of University Association for Science and Technology in Shaanxi under Grant 20210121, in part by the Scientific Research
%Program Funded by Shaanxi Provincial Education Department under Grant 21JK0914, in part by the  China Railway First Survey and Design Institute Group Co., LTD. Research Program under Grant 2022KY52ZD(ZNXT)-03.}
}
\markboth{}
{Shi\MakeLowercase{\textit{et al.}}:}
\maketitle
%focuses on the design of an optimal resource allocation scheme to maximize the energy efficiency (EE) for
\begin{abstract}
Ambient backscatter communications (AmBC), where a backscatter transmitter (BT) modulates and reflects ambient signals to a backscatter receiver (BR), have been deemed a low-power communication technology  for the Internet of Things. Previous work on symbol detection in AmBC assumed perfect time synchronization (TS), which is unrealistic in practice. The residual TS errors (RTSE) cause \emph{partial sample mismatch}, degrading symbol detection performance. To address this, we propose a new AmBC symbol detection framework that incorporates the BT's current and adjacent symbols, as well as channel coefficients. Using energy detector (ED) as a case study, we derive both exact and approximate bit error rate (BER) expressions. Our results show that the ED's BER performance degrades significantly under RTSE, with the symbol detection threshold optimized under the assumption of perfect TS. We then derive a closed-form expression for a near-optimal symbol detection threshold that minimizes BER under RTSE. To estimate the required parameters for the detection threshold, we propose a novel method  exploiting the attributes of the BR's received signal samples. The analytical results are verified by simulation results.
\end{abstract}
\IEEEpeerreviewmaketitle
\section{Introduction}
\IEEEPARstart{I}{n} the foreseeable future of the Internet of Things (IoT), we anticipate the widespread deployment of massively scaled, ultra-low complexity, and ultra-low power IoT devices that may rely on the ambient backscatter communications (AmBC)
\cite{9928079,9051982,10463656,9950349,chen2025high}. In AmBC, the backscatter transmitter (BT) emits information to the backscatter receiver (BR) by piggybacking its own modulated message onto incident  radio frequency (RF) signals \cite{9481248,9866050,10528254}. This approach simplifies circuitry and reduces power consumption of the BT, making it ideal for IoT applications. However, the symbol detection at the BR faces a challenge as the BR simultaneously receives ambient RF signals and BT-reflected signals, with the latter often being overshadowed by the former.

\subsection{Symbol Detection-Related Work}
The authors in \cite{7341107} derived a maximum likelihood (ML) detector and provided a closed-form expression of the symbol detection threshold that minimizes the bit error rate (BER). Although the ML detector achieves satisfactory performance, it requires prior knowledge of detection-required parameters, e.g., the channel coefficients and the transmit power of ambient RF signals. The authors in \cite{8007328} derived an energy detector (ED) and estimated detection-required parameters from received signal samples and pilot symbols. The authors in \cite{10269022} considered a reflecting surface (RS)-aided AmBC \textcolor{black}{system}, and presented accurate BER expressions of ED under an intelligent RS with ideal/non-ideal phase shifts and a dumb RS. The authors in \cite{10531099} proposed a p-norm based ED to improve the performance of ED. In addition to  ED, the authors in \cite{8865663} proposed a maximum-eigenvalue-based detector for a multi-antenna BR, where detection-required parameters were estimated by using pilot symbols. In \cite{9250656}, a deep transfer learning-based signal detection framework was proposed by employing convolutional neural networks to implicitly extract channel features and directly recover the BT's symbols.

To improve the symbol detection performance, coding methods were also used at the BT. The differential encoding was employed in \cite{7417704,7551180,10005249}. Specifically, a difference of energy based detector was proposed, and the corresponding near-optimal symbol detection threshold was estimated via making full use of the statistical characteristics of the received signal samples at the BR. Symbol detection has also been studied under the Manchester coding \cite{8329444}, the non-return-to-zero coding \cite{9242274} and the orthogonal space-time block coding \cite{9430725}. In \cite{8329444}, the Manchester coding and differential Manchester coding were adopted at the BT, and the corresponding semi-coherent Manchester and non-coherent Manchester detectors were developed. In \cite{9242274}, the probability density function (PDF) of the BR's received signal samples under the non-return-to-zero coding was derived and two non-coherent detectors were designed. In \cite{9430725}, the authors designed two coherent detectors and a non-coherent detector based on the orthogonal space-time block coding.
\subsection{Limitations of Existing Work}
We note that the existing works \cite{7341107,8007328,10531099,10269022,8865663,9250656,7417704,7551180,10005249,8329444,9242274,9430725} on the symbol detection in AmBC assumed perfect time synchronization (TS) between BT and BR. However, achieving perfect TS in practical AmBC systems is challenging. This is because implementing TS requires the BT to modulate TS sequences onto incident RF signals, which have gone through wireless channel fading, and reflect the modulated signals to the BR. As a result, the reflected signal is often overshadowed by the ambient RF signal, making perfect TS difficult to achieve. While several TS algorithms for AmBC have been proposed in \cite{dunna2021syncscatter,chi2020leveraging,10000657,10167801,8103807,10485514}, complete elimination of TS errors remains unattainable, and residual TS errors (RTSE) inevitably exist in practical systems. This RTSE leads to \emph{partial sample mismatch}, i.e., the BR's sampling interval for the BT's current symbol contains samples from adjacent symbols transmitted by the BT. However, AmBC symbol detection under RTSE has not yet been explored. In particular, the impacts of RTSE on symbol detection remain unclear, and how to mitigate the resulting performance degradation is an open issue that must be urgently addressed.

\subsection{Motivation and Contributions}
In this paper, we investigate an AmBC \textcolor{black}{system}, which consists of one ambient RF source (S) and its associated receiver, one BT, and one BR, under RTSE between BT and BR. Our goal is to examine the impacts of RTSE on BER of AmBC symbol detector, and propose a symbol detection method that works well under RTSE by exploiting the attributes of all received samples. \textcolor{black}{The ED is used as a case study due to its simplicity and practical applicability in scenarios with limited computational resources and hardware budgets, as well as its robustness against frequency offsets and phase fluctuations.} The main contributions are summarized as follows.
\begin{itemize}
\item We propose a new symbol detection framework  for AmBC under RTSE, which includes all the eight possible cases of partial sample mismatch due to  RTSE.
\item We derive the exact BER of ED into a single integral form. To reduce the computational complexity in assessing  the BER achieved by the ED, we employ the Lyapunov central limit theorem (CLT) to derive an approximate yet concise expression of the BER. It reveals that if the symbol detection threshold obtained assuming  perfect TS is directly used in practical AmBC under RTSE, a serious degradation  of BER cannot be avoided.
\item  Leveraging approximate BER expression, we derive a closed-form expression for the near-optimal symbol detection threshold that minimizes BER of ED under RTSE.
\item The near-optimal symbol detection threshold requires prior knowledge of the channel coefficients, the RTSE, the noise power, and the transmission power of S, which is unknown by the BR in practice. To address this, we propose a novel method to estimate above parameters by exploiting the attributes of the BR's received signal samples. Different the existing works, e.g., \cite{8007328}, our method can work in the presence of RTSE.
\item In practice, S may be a phase shift keying (PSK) signal rather than a complex Gaussian signal. To extend the applicability of our symbol detection method, we also analyze ED performance with a PSK signal under RTSE.
\item Computer simulations show accuracy of our obtained symbol detection threshold, and confirm that using our proposed detection threshold, the BER can be significantly reduced compared to that using the threshold obtained by existing work \cite{8007328}.
\end{itemize}

%The rest of this paper is organized as follows. In Section II, we introduce the symbol detection model under perfect TS and then propose the symbol detection model under RTSE. In Section III, we analyze the influence of RTSE on BER. In Section IV, we obtain the near-optimal symbol detection threshold under RTSE, which can be estimated by our proposed   method without any prior knowledge. Numerical results are presented in Section V, while conclusions are drawn in Section VI.

The main notations used in this paper are listed below.
$\mathbb{E} \left[ x \right]$, $\mathbb{D} \left[ x \right]$, and $\left| x \right|$ denote the expectation, the variance, and the absolute value of $x$, respectively. \textcolor{black}{$\mathbb{N}(a,b)$ and $\mathbb{CN}(a,b)$ denote the Gaussian distribution and the complex Gaussian distribution with mean $a$ and variance $b$, respectively.}

\section{Symbol Detection Model}
\begin{figure}
  \centering
  \includegraphics[scale=0.5]{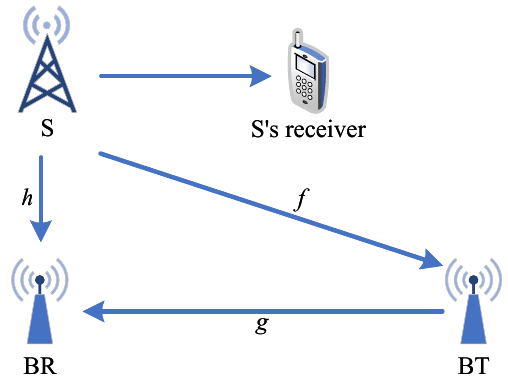}\\
  \caption{AmBC system.}\label{fig1}
  \vspace{-15pt}
\end{figure}

We consider an AmBC system with an S, a BR, and a BT, each equipped with a single antenna\footnote{The primary objective of this work is to investigate the impact of RTSE on symbol detection and minimize its effects. To streamline the analysis and ensure a focused study on RTSE, we opt for a single-antenna BT and BR, thereby reducing the complexity of the analysis. It is important to note that our proposed method is also applicable to multi-antenna BR systems or multi-antenna Ss. In the case of the former, our approach can be applied independently to each antenna, with the multi-antenna detection results integrated using fusion criteria, such as OR/AND rules. For the latter, the proposed method can be directly applied, as the detection threshold estimation method does not require the prior
knowledge of the S's transmit power (see Section IV), and the multiple antennas at the S only influence the power in specific directions.}, as shown in Fig. \ref{fig1}. In this system, S transmits information symbol $s(n)$ to its associated receiver, while the BT modulates and reflects the incident signal to carry its own binary information symbols $B(k)$ to BR. Let $h$, $f$ and $g$ represent the complex channel coefficients of the S$-$BR link, the S$-$BT link, and the BT$-$BR link, respectively. We assume that all channels follow independent frequency flat block fading, i.e., all channel coefficients do not change within a transmission block. Each transmission block contains $K$ symbol periods of the BT.
\subsection{Symbol Detection  Under Perfect TS}
\setcounter{equation}{5}
\begin{figure*}[t]
\normalsize
\begin{align}\label{6}
\textcolor{black}{
y_{k,\Delta n < 0}^{{\rm{ip}}}\left( n \right) = \left\{ {\begin{array}{*{20}{l}}
{{{\cal Y}_h}\left( n \right),n = n_{{\rm{start}}}^{\left( k \right)}, \cdots ,n_{{\mathop{\rm end}\nolimits} }^{\left( k \right)},}&{{\rm{if}}\;B\left( {k - 1} \right) = 0,B\left( k \right) = 0}\\
{\left. {\begin{array}{*{20}{l}}
{{{\cal Y}_\mu }\left( n \right),n = n_{{\rm{start}}}^{\left( k \right)}, \cdots ,n_a^{\left( k \right)}}\\
{{{\cal Y}_h}\left( n \right),n = n_a^{\left( k \right)} + 1, \cdots ,n_{{\mathop{\rm end}\nolimits} }^{\left( k \right)}}
\end{array}} \right\},}&{{\rm{if}}\;B\left( {k - 1} \right) = 1,B\left( k \right) = 0}\\
{{{\cal Y}_\mu }\left( n \right),n = n_{{\rm{start}}}^{\left( k \right)}, \cdots ,n_{{\mathop{\rm end}\nolimits} }^{\left( k \right)},}&{{\rm{if}}\;B\left( {k - 1} \right) = 1,B\left( k \right) = 1}\\
{\left. {\begin{array}{*{20}{l}}
{{{\cal Y}_h}\left( n \right),n = n_{{\rm{start}}}^{\left( k \right)}, \cdots ,n_a^{\left( k \right)}}\\
{{{\cal Y}_\mu }\left( n \right),n = n_a^{\left( k \right)} + 1, \cdots ,n_{{\mathop{\rm end}\nolimits} }^{\left( k \right)}}
\end{array}} \right\},}&{{\rm{if}}\;B\left( {k - 1} \right) = 0,B\left( k \right) = 1}
\end{array}} \right.,}
\end{align}
\vspace{-15pt}
\hrulefill
\end{figure*}
Since BT's symbol rate is much slower than that of S, we assume that each BT's symbol $B(k)$ remains unchanged within $N$ consecutive S' samples  $s(n)$ \cite{7341107,8007328,10531099,10269022,8865663,9250656,7417704,7551180,10005249}. Then in the case of perfect TS  between  BT and  BR, the $n$\textcolor{black}{-}th sample of the $k$\textcolor{black}{-}th symbol received by the BR is written as \footnote{Here we omit the thermal noise for the reason as detailed in \cite{7551180}. It is worth noting that our proposed symbol detection method is also applicable if this noise is considered. This is because our proposed method does not require knowing the  power of $\omega (n)$.}\cite{8007328}
\setcounter{equation}{0}
\begin{align}\label{1}
y_k^{\rm{p}}(n) = hs(n)+ \eta fgB(k)s(n) + \omega (n),
\end{align}
where $n=(k - 1)N + 1,(k - 1)N + 2, \cdots ,kN$, $\eta$ is the complex attenuation of the signal inside the BT\footnote{$\eta$ determines the amount of energy harvested by the BT, and the harvested energy is utilized to compensate for circuitry power consumption of the BT.}, \textcolor{black}{$s\left( n \right) \sim \mathbb{CN}\left( {0,{P_s}} \right)$} with the power $P_s$, and \textcolor{black}{$\omega \left( n \right) \sim \mathbb{CN}\left( {0,{N_\omega }} \right)$} is an additive white complex Gaussian noise with the power ${N_\omega }$. %{\color{blue}Here we omit the thermal noise\footnote{{\color{blue}It is worth noting that our proposed symbol detection method is also applicable if this noise is considered. This is because our proposed method does not require the  power of $\omega (n)$.}} and the detailed reasons can be referred to \cite{7551180}.}
Considering an on-off keying (OOK) modulation at the BT \textcolor{black}{\footnote{\textcolor{black}{In this work, we consider the BT with OOK modulation, where $B(k)$ takes values of `0' and `1'. The symbol $B(k)$ is transmitted by changing the reflection coefficient ${\Gamma _i}$, i.e., the reflection coefficient is equal to the value of $B(k)$. Therefore, ${\Gamma _i} \in \left\{ {0,1} \right\}$ is the reflection coefficient, which is the same as that in \cite{liu2013ambient,8368232,7551180}.}}}, we rewrite $y_k^{\rm{p}}(n)$  as \textcolor{black}{\footnote{The backscattered signal at BR is composed of two different components: structural mode and antenna mode scattering \cite{6685977}. By considering the structural mode scattering as part of the interference from
the direct link and the antenna mode scattering as part of the backscatter link \cite{8907447}, we obtain (2).}}
\setcounter{equation}{1}
\begin{align}\label{2}
y_k^{\rm{p}}(n) = \left\{ {\begin{array}{*{20}{c}}
{hs(n) + \omega (n),\;{\rm{if}}\;B\left( k \right) = 0}\\
{\mu s(n) + \omega (n),\;{\rm{if}}\;B\left( k \right) = 1}
\end{array}} \right.,
\end{align}
where
$\mu=h+\eta fg$.
\setcounter{equation}{6}
\begin{figure*}[t]
\normalsize
\begin{align}\label{7}
\textcolor{black}{
y_{k,\Delta n > 0}^{{\rm{ip}}}\left( n \right) = \left\{ {\begin{array}{*{20}{l}}
{{{\cal Y}_h}\left( n \right),n = n_{{\rm{start}}}^{\left( k \right)}, \cdots ,n_{{\mathop{\rm end}\nolimits} }^{\left( k \right)},}&{{\rm{if}}\;B\left( k \right) = 0,B\left( {k + 1} \right) = 0}\\
{\left. {\begin{array}{*{20}{l}}
{{{\cal Y}_h}\left( n \right),n = n_{{\rm{start}}}^{\left( k \right)}, \cdots ,n_{{\mathop{\rm end}\nolimits} }^{\left( k \right)} - {n_a}}\\
{{{\cal Y}_\mu }\left( n \right),n = n_{{\mathop{\rm end}\nolimits} }^{\left( k \right)} - {n_a} + 1, \cdots ,n_{{\mathop{\rm end}\nolimits} }^{\left( k \right)}}
\end{array}} \right\},}&{{\rm{if}}\;B\left( k \right) = 0,B\left( {k + 1} \right) = 1}\\
{{{\cal Y}_\mu }\left( n \right),n = n_{{\rm{start}}}^{\left( k \right)}, \cdots ,n_{{\mathop{\rm end}\nolimits} }^{\left( k \right)},}&{{\rm{if}}\;B\left( k \right) = 1,B\left( {k + 1} \right) = 1}\\
{\left. {\begin{array}{*{20}{l}}
{{{\cal Y}_\mu }\left( n \right),n = n_{{\rm{start}}}^{\left( k \right)}, \cdots ,n_{{\mathop{\rm end}\nolimits} }^{\left( k \right)} - {n_a}}\\
{{{\cal Y}_h}\left( n \right),n = n_{{\mathop{\rm end}\nolimits} }^{\left( k \right)} - {n_a} + 1, \cdots ,n_{{\mathop{\rm end}\nolimits} }^{\left( k \right)}}
\end{array}} \right\},}&{{\rm{if}}\;B\left( k \right) = 1,B\left( {k + 1} \right) = 0}
\end{array}} \right.,}
\end{align}
\vspace{-17pt}
\hrulefill
\end{figure*}

The BR adopts ED to estimate  $B(k)$, then  following [\cite{8007328}, eq.(23)], we have
\setcounter{equation}{2}
\begin{align}\label{3}
{\begin{array}{*{20}{c}}
{\hat B\left( k \right) = 0,\left\{ {\begin{array}{*{20}{c}}
{{\rm{if}}\;\Gamma _k^{\rm{p}} \ge \gamma _{{\rm{th}}}^{\rm{p}},\sigma _0^2 > \sigma _1^2}\\
{{\rm{if}}\;\Gamma _k^{\rm{p}} < \gamma _{{\rm{th}}}^{\rm{p}},\sigma _0^2 \le \sigma _1^2}
\end{array}} \right.}\\
{\hat B\left( k \right) = 1,\left\{ {\begin{array}{*{20}{c}}
{{\rm{if}}\;\Gamma _k^{\rm{p}} < \gamma _{{\rm{th}}}^{\rm{p}},\sigma _0^2 > \sigma _1^2}\\
{{\rm{if}}\;\Gamma _k^{\rm{p}} \ge \gamma _{{\rm{th}}}^{\rm{p}},\sigma _0^2 \le \sigma _1^2}
\end{array}} \right.}
\end{array}},
\end{align}
where $\Gamma _k^{\rm{p}} = \sum\limits_{n = (k - 1)N + 1}^{kN} {{{\left| {y_k^{\rm{p}}(n)} \right|}^2}} $ is the total energy of the $N$ consecutive samples corresponding to $B(k)$, ${\gamma _{{\rm{th}}}^{{\rm{p}}}}$ is the symbol detection threshold under perfect TS, $\sigma _0^2 = {\left| h \right|^2}{P_s} + {N_\omega }$, $\sigma _1^2 = {\left| \mu  \right|^2}{P_s} + {N_\omega }$, and ${\hat B\left( k \right)}$ is the estimated value of $B(k)$. %The details can be refer to Section III-B in \cite{8007328}.

By assuming $\Pr \left( {B\left( k \right) = 0} \right) = \Pr \left( {B\left( k \right) = 1} \right)=\frac{1}{2}$, the BER and the optimal symbol detection threshold under perfect TS  can be, respectively, derived as \cite{8007328}
\begin{align} \label{4}
P_{{\rm{BER}}}^{\rm{p}} &= \frac{1}{2}Q\left( {\frac{{\gamma _{{\rm{th}}}^{\rm{p}} - N\sigma _{\min }^2}}{{\sqrt N \sigma _{\min }^2}}} \right) \!+\! \frac{1}{2}Q\left( {\frac{{N\sigma _{\max }^2 - \gamma _{{\rm{th}}}^{\rm{p}}}}{{\sqrt N \sigma _{\max }^2}}} \right),
\end{align}
\begin{align}\label{5}
\gamma _{{\rm{th,opt}}}^{\rm{p}} = \frac{{2N\sigma _0^2\sigma _1^2}}{{\sigma _0^2 + \sigma _1^2}},
\end{align}
where $\sigma _{\max }^2 = \max \left\{ {\sigma _0^2,\sigma _1^2} \right\}$, $\sigma _{\min }^2 = \min \left\{ {\sigma _0^2,\sigma _1^2} \right\}$, and $Q\left( x \right) = \frac{1}{{\sqrt {2\pi } }}\int_x^\infty  {{e^{ - \frac{{{t^2}}}{2}}}} dt$ is the $Q$ function.

{\emph{Remark 1.}} \textcolor{black}{The optimal symbol detection threshold \eqref{5} is achieved when the value of $N$ is sufficiently large.} When the optimal symbol detection threshold is used, $\Pr \left( {\hat B\left( k \right) = 1|B\left( k \right) = 0} \right) = \Pr \left( {\hat B\left( k \right) = 0|B\left( k \right) = 1} \right)$ can be achieved, which is generally referred to as achieving balanced BER \cite{6409345}.

\setcounter{equation}{8}
\begin{figure*}[t]
\normalsize
\begin{align}\label{9}
\Gamma _k^{{\rm{ip}}} = \Gamma _{k|i,j}^{{\rm{ip}}} = \left\{ {\begin{array}{*{20}{c}}
{\sum\limits_{n = (k - 1)N + 1}^{(k - 1)N + {n_a}} {{{\left| {y_{k - 1}^{{\rm{ip}}}\left( n \right)|B\left( {k - 1} \right) = i} \right|}^2}}  + \sum\limits_{n = (k - 1)N + {n_a} + 1}^{kN} {{{\left| {y_k^{{\rm{ip}}}\left( n \right)|B\left( k \right) = j} \right|}^2}} ,}&{{\rm{if}}\;\Delta n < 0}\\
{\sum\limits_{n = (k - 1)N + 1}^{kN - {n_a}} {{{\left| {y_k^{{\rm{ip}}}\left( n \right)|B\left( k \right) = j} \right|}^2}}  + \sum\limits_{n = kN - {n_a} + 1}^{kN} {{{\left| {y_{k + 1}^{{\rm{ip}}}\left( n \right)|B\left( {k + 1} \right) = i} \right|}^2}} ,}&{{\rm{if}}\;\Delta n > 0}
\end{array}} \right.,
\end{align}
\vspace{-17pt}
\hrulefill
\end{figure*}

\begin{figure*}[t]
\normalsize
\begin{align}\label{10}
P_{{\rm{BER}}}^{{\rm{ip}}} = \left\{ {\begin{array}{*{20}{c}}
{\frac{1}{4}\int\limits_{ - \infty }^{\gamma _{{\rm{th}}}^{{\rm{ip}}}} {{f_{\Gamma _{k|0,0}^{{\rm{ip}}}}}\left( z \right)dz}  + \frac{1}{4}\int\limits_{ - \infty }^{\gamma _{{\rm{th}}}^{{\rm{ip}}}} {{f_{\Gamma _{k|1,0}^{{\rm{ip}}}}}\left( z \right)dz}  + \frac{1}{4}\int\limits_{\gamma _{{\rm{th}}}^{{\rm{ip}}}}^\infty  {{f_{\Gamma _{k|1,1}^{{\rm{ip}}}}}\left( z \right)dz}  + \frac{1}{4}\int\limits_{\gamma _{{\rm{th}}}^{{\rm{ip}}}}^\infty  {{f_{\Gamma _{k|0,1}^{{\rm{ip}}}}}\left( z \right)dz} ,{\rm{if}}\;\sigma _0^2 > \sigma _1^2}\\
{\frac{1}{4}\int\limits_{\gamma _{{\rm{th}}}^{{\rm{ip}}}}^\infty  {{f_{\Gamma _{k|0,0}^{{\rm{ip}}}}}\left( z \right)dz}  + \frac{1}{4}\int\limits_{\gamma _{{\rm{th}}}^{{\rm{ip}}}}^\infty  {{f_{\Gamma _{k|1,0}^{{\rm{ip}}}}}\left( z \right)dz}  + \frac{1}{4}\int\limits_{ - \infty }^{\gamma _{{\rm{th}}}^{{\rm{ip}}}} {{f_{\Gamma _{k|1,1}^{{\rm{ip}}}}}\left( z \right)dz}  + \frac{1}{4}\int\limits_{ - \infty }^{\gamma _{{\rm{th}}}^{{\rm{ip}}}} {{f_{\Gamma _{k|0,1}^{{\rm{ip}}}}}\left( z \right)dz} ,{\rm{if}}\;\sigma _0^2 \le \sigma _1^2}
\end{array}} \right.,
\end{align}
\vspace{-20pt}
\hrulefill
\end{figure*}

\subsection{Symbol Detection  Under RTSE}
\begin{figure}
  \centering
  \includegraphics[width=0.45\textwidth]{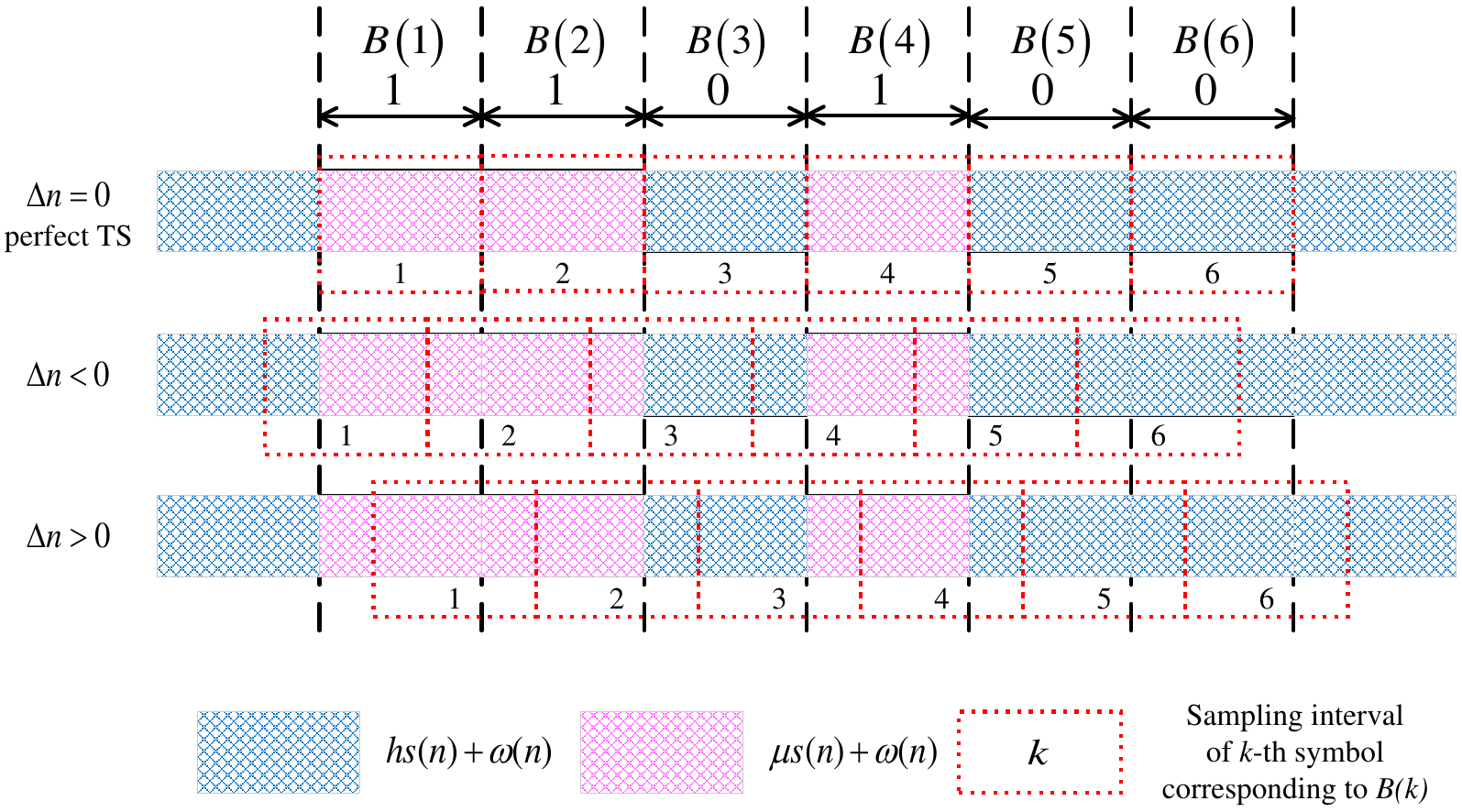}\\
  \caption{BR's received signal under RTSE.}\label{fig2}
\end{figure}

Under RTSE, the real arrival time of BT's signal, denoted by $n$, is not exactly estimated by the BR. In such a case, the estimated arrival time of BT's signal, denoted by $\widehat{n}$, is smaller (or larger) than the real one $n$,  resulting in a {\emph{partial sample mismatch}}, i.e., the sampling interval corresponding to $B(k)$ contains the samples corresponding to its adjacent symbols, i.e, $B(k-1)$ or $B(k+1)$. This can be illustrated  by Fig. \ref{fig2}, where BT modulates and reflects symbols ``110100'', and $\Delta n= \widehat{n}-n$. Taking $\Delta n  < 0$ as an example, the sampling interval corresponding to $B(4)=1$ contains the samples corresponding to $B(3)=0$. In this case, the resulting $\Pr \left( {\hat B\left( 4 \right) = 0|B\left( 4 \right) = 1} \right)$ is larger than that under perfect TS (i.e., $\Delta n = 0$), indicating the degradation of detection performance. We also note that although the sampling interval corresponding to $B(2)=1$ contains the samples corresponding to $B(1)=1$, $\Pr \left( {\hat B\left( 2 \right) = 0|B\left( 2 \right) = 1} \right)$ equals that under perfect TS (i.e., $\Delta n = 0$). In this case, the partial samples mismatch has no impact on estimating $B(2)$. In summary, there exists a partial samples mismatch under RTSE and whether or not the partial samples mismatch degrades the detection performance depends on $\Delta n$ and the relationship between adjacent symbols. The above observations make the symbol detector under RTSE quite different from the previous works that assume perfect TS \cite{7341107,8007328,10531099,10269022,8865663,9250656,7417704,7551180,10005249,8329444,9242274,9430725} and it is urgent to revisit the symbol detection under RTSE.

Towards this end, the symbol detection model under RTSE is given by \eqref{6} and \eqref{7}, as shown at the top of this page and next page, respectively, where the superscript `ip' denotes the RTSE, ${{\cal Y}_h}\left( n \right) = hs(n) + \omega (n)$, ${{\cal Y}_\mu }\left( n \right) = \mu s(n) + \omega (n)$, $n_{{\rm{start}}}^{\left( k \right)} = \left( {k - 1} \right)N + 1$, $n_{{\mathop{\rm end}\nolimits} }^{\left( k \right)} = kN$, $n_a^{\left( k \right)} = \left( {k - 1} \right)N + {n_a}$, and ${n_a} = \left| {\Delta n} \right|$ reflects the level of RTSE, whose exact value is unknown but can be estimated by the BR
using the method proposed in Section IV-B.

%\textcolor{black}{{\emph{Remark 2.}} We use the total energy of $N$ consecutive samples to recover BT's symbols (as detailed in the next section) and the detection of variation edges is not required. This approach avoids the RTSE fluctuations  across the backscatter symbols transmitted within a transmission block. Hence, we assume a constant RTSE within a transmission block. Since the TS is performed at the beginning of each transmission block, the RTSE may vary between adjacent transmission blocks.}

{\emph{Remark 2.}} In practical AmBC, since the TS algorithm is implemented before symbol detection, then only a small RTSE exists. For example, the recent experimental findings  demonstrate that in an LTE backscatter system, the RTSE can be up to 120 microseconds (see Fig. 16 in \cite{10167801}). In this case, with a transmission rate of 1 Kbps using BPSK modulation, the RTSE ratio, which is defined as the RTSE divided by the duration of each BT symbol, can peak at 12\%. Keeping this in mind, we assume $\frac{{{n_a}}}{N} < 50\%$ and this assumption will be used to estimate the symbol detection threshold in Section IV-B.

By comparing \eqref{6} and \eqref{7} with \eqref{2}, it can be seen that $y_k^{{\rm{ip}}}\left( n \right) = y_k^{\rm{p}}\left( n \right)$ for $n_a=0$, i.e., the perfect TS is a special case under RTSE. However, for ${n_a} \ne 0$, the symbol detection model under RTSE is more complex than that with perfect TS. In this case, if the BR directly employs ED with the optimal symbol detection threshold \eqref{5}, the detection performance will degrade, which will be studied in the next section.
\section{Impacts of RTSE on ED}
In this section, we derive the BER as a function of the symbol detection threshold and the RTSE, based on which we elaborate the impacts of RTSE on ED. Towards this end, we first present the decision criterion under RTSE \textcolor{blue}{\footnote{The  pilot symbol, transmitted during the TS  phase, enables the determination of whether  $\sigma _0^2$ or $\sigma _1^2$ is larger.}}, given by
\setcounter{equation}{7}
\begin{align}\label{8}
{\begin{array}{*{20}{c}}
{\hat B\left( k \right) = 0,\left\{ {\begin{array}{*{20}{c}}
{{\rm{if}}\;\Gamma _k^{{\rm{ip}}} \ge \gamma _{{\rm{th}}}^{{\rm{ip}}},\sigma _0^2 > \sigma _1^2}\\
{{\rm{if}}\;\Gamma _k^{{\rm{ip}}} < \gamma _{{\rm{th}}}^{{\rm{ip}}},\sigma _0^2 \le \sigma _1^2}
\end{array}} \right.}\\
{\hat B\left( k \right) = 1,\left\{ {\begin{array}{*{20}{c}}
{{\rm{if}}\;\Gamma _k^{{\rm{ip}}} < \gamma _{{\rm{th}}}^{{\rm{ip}}},\sigma _0^2 > \sigma _1^2}\\
{{\rm{if}}\;\Gamma _k^{{\rm{ip}}} \ge \gamma _{{\rm{th}}}^{{\rm{ip}}},\sigma _0^2 \le \sigma _1^2}
\end{array}} \right.}
\end{array}},
\end{align}
where ${\gamma _{{\rm{th}}}^{{\rm{ip}}}}$ is the symbol detection threshold under RTSE, and its near-optimal value to minimize BER is derived in Section IV-A, and $\Gamma _k^{{\rm{ip}}}$ is expressed as \eqref{9}, as shown at the top of this page, which denotes the total energy of $N$ consecutive samples corresponding to the cases $\left\{ {B\left( {k - 1} \right) = i,B\left( k \right) = j, i,j \in \left\{ {0,1} \right\}} \right\}$ when $\Delta n < 0$ (or $\left\{ {B\left( {k} \right) = j,B\left( {k+1} \right) = i} \right\}$ when $\Delta n > 0$).

\textcolor{black}{\emph{Proof.} The rationale behind the effectiveness of decision criterion \eqref{8}.}

\textcolor{black}{Please refer to Appendix A. \hfill {$\blacksquare $}}

Using \eqref{8} and the fact that  $\Pr \left( {B\left( k \right) = 0} \right) = \Pr \left( {B\left( k \right) = 1} \right)$ holds in most practical wireless communications \cite{6926805}, we can obtain the exact BER of the ED under RTSE, as summarized in Theorem 1.

{\textbf{Theorem 1.}} For a given symbol detection threshold ${\gamma _{{\rm{th}}}^{{\rm{ip}}}}$ and a RTSE $n_a$, the BER achieved by the ED is written as \eqref{10}, as shown at the top of this page, where ${{f_{\Gamma _{k|i,j}^{{\rm{ip}}}}}\left( z \right)}$ denotes the PDF of the random variables $\Gamma _{k|i,j}^{{\rm{ip}}}$, and ${{f_{\Gamma _{k|0,0}^{{\rm{ip}}}}}\left( z \right)}$, ${{f_{\Gamma _{k|1,1}^{{\rm{ip}}}}}\left( z \right)}$, ${{f_{\Gamma _{k|0,1}^{{\rm{ip}}}}}\left( z \right)}$ and ${{f_{\Gamma _{k|1,0}^{{\rm{ip}}}}}\left( z \right)}$ are given by \eqref{B2}, \eqref{B3}, \eqref{B9} and \eqref{B10}, respectively.

{\emph{Proof.}} Please refer to Appendix B.   \hfill {$\blacksquare $}

It can be seen from \eqref{10} that the expression of BER is too complex to assess the BER achieved by the ED. This  is mainly because $\Gamma _{k|i,j}^{{\rm{ip}}}$ are independent but not identically distributed (i.n.i.d),  and  thus difficult to directly combine. Fortunately, the Lyapunov CLT \cite{billingsley1995probability} provides a solution to this issue, and is introduced as follows.

{\bf{Lyapunov CLT.}} Let ${X_1},{X_2}, \cdots ,{X_N}, \cdots $ be independent random variable sequences with mathematical expectation $\mathbb{E} \left[ {{X_n}} \right] = {\mu _n}$ and variance $\mathbb{D} \left[ {{X_n}} \right] = \sigma _n^2 > 0$, where $n = 1,2, \cdots, N, \cdots $. Define ${s_N} = \sqrt {\sum\limits_{n = 1}^N {\sigma _n^2} } $. If there exists $\delta  > 0$ such that the Lyapunov condition $\mathop {\lim }\limits_{N \to \infty }  \frac{{\sum\limits_{n = 1}^N  \mathbb{E}\left[ {{{\left| {{X_n} - {\mu _n}} \right|}^{2 + \delta }}} \right]}}{{{{\left( {{s_N}} \right)}^{2 + \delta }}}} = 0$ \cite{billingsley1995probability} holds, then we have $\mathop {\lim }\limits_{N \to \infty } P\left\{ {\frac{{\sum\limits_{n = 1}^N {{X_n}}  - \sum\limits_{n = 1}^N {{\mu _n}} }}{{{s_N}}} \le x} \right\} = \int_{ - \infty }^x {\frac{1}{{\sqrt {2\pi } }}} {e^{ - \frac{{{t^2}}}{2}}}dt$.  \hfill {$\blacksquare $}

In what follows, we approximate a closed-form expression of the BER by using Lyapunov CLT. To this end, Lemma 1 is provided.

\color{black}
\begin{figure*}[t]
\normalsize
\begin{align}\label{16}
\textcolor{black}{
\left\{ {\begin{array}{*{20}{c}}
{\Pr \left( {\Gamma _{k|i,0}^{{\rm{ip}}} \ge \gamma _{{\rm{th,opt}}}^{{\rm{ip}}}|B\left( k \right) = 0,{{\tilde B}_{\Delta n}}\left( k \right) = i} \right) = \Pr \left( {\Gamma _{k|i,1}^{{\rm{ip}}} < \gamma _{{\rm{th,opt}}}^{{\rm{ip}}}|B\left( k \right) = 1,{{\tilde B}_{\Delta n}}\left( k \right) = i} \right),{\rm{if}}\;\sigma _0^2 \le \sigma _1^2}\\
{\Pr \left( {\Gamma _{k|i,0}^{{\rm{ip}}} < \gamma _{{\rm{th,opt}}}^{{\rm{ip}}}|B\left( k \right) = 0,{{\tilde B}_{\Delta n}}\left( k \right) = i} \right) = \Pr \left( {\Gamma _{k|i,1}^{{\rm{ip}}} \ge \gamma _{{\rm{th,opt}}}^{{\rm{ip}}}|B\left( k \right) = 1,{{\tilde B}_{\Delta n}}\left( k \right) = i} \right),{\rm{if}}\;\sigma _0^2 > \sigma _1^2}
\end{array}} \right.,}\tag*{\textcolor{black}{(16)}}
\end{align}
\vspace{-20pt}
\hrulefill
\end{figure*}

{\textbf{Lemma 1.}} The test statistic $\Gamma _{k|i,j}^{{\rm{ip}}}$ satisfies the Lyapunov condition, given by
\vspace{-5pt}
\setcounter{equation}{10}
\begin{align}\label{11}
\mathop {\lim }\limits_{N \to \infty } \!\!\frac{{\sum\limits_{n \!=\! \left( {k \!-\! 1} \right)N + 1}^{kN}\!\!\!\! {\mathbb{E}\left[ {{{\left| {{{\left| {y_k^{{\rm{ip}}}\left( n \right)} \right|}^2} \!\!\!-\! \mathbb{E}\left[ {{{\left| {y_k^{{\rm{ip}}}\left( n \right)} \right|}^2}} \right]} \right|}^{2 + \delta }}} \right]} }}{{{{\left( {{s_{N,k}}} \right)}^{2 + \delta }}}} \!=\! 0,
\end{align}
\vspace{-5pt}
where ${s_{N,k}} = \sqrt {\sum\limits_{n = (k - 1)N + 1}^{kN} {\mathbb{D}\left[ {{{\left| {y_k^{{\rm{ip}}}\left( n \right)} \right|}^2}} \right]} } $, $\delta$ is a positive constant chosen at random.

{\emph{Proof.}} Please refer to Appendix C.   \hfill {$\blacksquare $}

Based on Lemma 1 and the Lyapunov CLT, we can approximate $\Gamma _{k|i,j}^{{\rm{ip}}}$ as a Gaussian distribution, i.e., $\Gamma _{k|i,j}^{{\rm{ip}}} \sim \mathbb{N}\left( {{\mu _{i,j}},{\varsigma _{i,j}}} \right)$. Then, we obtain the PDF of $\Gamma _{k|i,j}^{{\rm{ip}}}$ as
\begin{align}\label{12}
{f_{\Gamma _{k|i,j}^{{\rm{ip}}}}}\left( z \right) = \frac{1}{{\sqrt {2\pi {\varsigma _{i,j}}} }}\exp \left[ { - \frac{{{{\left( {z - {\mu _{i,j}}} \right)}^2}}}{{2{\varsigma _{i,j}}}}} \right],
\end{align}
where $i \in \left\{ {0,1} \right\}$, $j \in \left\{ {0,1} \right\}$, ${{\mu _{0,0}} = N\sigma _0^2}$, ${{\varsigma _{0,0}} = N\sigma _0^4}$, ${{\mu _{1,0}} = {n_a}\sigma _1^2 + \left( {N - {n_a}} \right)\sigma _0^2}$, ${{\varsigma _{1,0}} = {n_a}\sigma _1^4 + \left( {N - {n_a}} \right)\sigma _0^4}$, ${{\mu _{1,1}} = N\sigma _1^2}$, ${{\varsigma _{1,1}} = N\sigma _1^4}$, ${{\mu _{0,1}} = {n_a}\sigma _0^2 + \left( {N - {n_a}} \right)\sigma _1^2}$ and ${{\varsigma _{0,1}} = {n_a}\sigma _0^4 + \left( {N - {n_a}} \right)\sigma _1^4}$.

Using \eqref{12} and \eqref{8}, we can obtain the BER of ED under RTSE, as summarized in Theorem 2.

{\textbf{Theorem 2.}} Given the symbol detection threshold ${\gamma _{{\rm{th}}}^{{\rm{ip}}}}$ and the RTSE $n_a$, the BER achieved by the ED can be approximated as
\vspace{-5pt}
\begin{align}\label{13} \notag
P_{{\rm{BER}}}^{{\rm{ip}}} &\simeq \frac{1}{4}Q\left( {\frac{{\gamma _{{\rm{th}}}^{{\rm{ip}}} - N\sigma _{\min }^2}}{{\sqrt N \sigma _{\min }^2}}} \right) + \frac{1}{4}Q\left( {\frac{{N\sigma _{\max }^2 - \gamma _{{\rm{th}}}^{{\rm{ip}}}}}{{\sqrt N \sigma _{\max }^2}}} \right)\\  \notag
&+ \frac{1}{4}Q\left( {\frac{{\gamma _{{\rm{th}}}^{{\rm{ip}}} - \left( {{n_a}\sigma _{\max }^2 + \left( {N - {n_a}} \right)\sigma _{\min }^2} \right)}}{{\sqrt {{n_a}\sigma _{\max }^4 + \left( {N - {n_a}} \right)\sigma _{\min }^4} }}} \right)\\  \notag
&+ \frac{1}{4}Q\left( {\frac{{\left( {{n_a}\sigma _{\min }^2 + \left( {N - {n_a}} \right)\sigma _{\max }^2} \right) - \gamma _{{\rm{th}}}^{{\rm{ip}}}}}{{\sqrt {{n_a}\sigma _{\min }^4 + \left( {N - {n_a}} \right)\sigma _{\max }^4} }}} \right). \tag{13}
\end{align}
\vspace{-5pt}

{\emph{Proof.}} Please refer to Appendix D.   \hfill {$\blacksquare $}

{\emph{Remark 3.}} The derived result \eqref{13} serves the following two purposes. Firstly, it assesses the achievable BER under any given symbol detection threshold and RTSE, thereby circumventing the need for numerous Monte Carlo simulation experiments. Secondly, it can be used to demonstrate the necessity to consider the RTSE in practical AmBC via the difference between the BER obtained by substituting \eqref{5} into \eqref{13} and the expected one obtained by substituting \eqref{5} into \eqref{4}. If the difference is tiny, we can say the impacts of RTSE can be negligible; otherwise, it is vital to take the RTSE into account for symbol detection. This question is answered in what follows.

The difference of BER between RTSE and perfect TS is given by
\vspace{-5pt}
\begin{align}\label{14}\notag
P_{{\rm{BER}}}^{\rm{d}} &= P_{{\rm{BER}}}^{{\rm{ip}}} - P_{{\rm{BER}}}^{\rm{p}}\\ \notag
&= \frac{1}{4}Q\left( {\frac{{\gamma _{{\rm{th,opt}}}^{\rm{p}} - \left( {{n_a}\sigma _{\max }^2 + \left( {N - {n_a}} \right)\sigma _{\min }^2} \right)}}{{\sqrt {{n_a}\sigma _{\max }^4 + \left( {N - {n_a}} \right)\sigma _{\min }^4} }}} \right)\\ \notag
&- \frac{1}{4}Q\left( {\frac{{\gamma _{{\rm{th,opt}}}^{\rm{p}} \!\!-\! N\sigma _{\min }^2}}{{\sqrt N \sigma _{\min }^2}}} \right) \!- \!\frac{1}{4}Q\left( {\frac{{N\sigma _{\max }^2\!\! -\! \gamma _{{\rm{th,opt}}}^{\rm{p}}}}{{\sqrt N \sigma _{\max }^2}}} \right)\\ \notag
&+ \frac{1}{4}Q\left( {\frac{{\left( {{n_a}\sigma _{\min }^2 \!+\! \left( {N \!-\! {n_a}} \right)\sigma _{\max }^2} \right)\! - \!\gamma _{{\rm{th,opt}}}^{\rm{p}}}}{{\sqrt {{n_a}\sigma _{\min }^4 + \left( {N - {n_a}} \right)\sigma _{\max }^4} }}} \right),\tag{14}
\end{align}
where the superscript `d' denotes the difference of the BER.

\begin{figure*}[t]
\normalsize
\begin{align}\label{20}
\textcolor{black}{
\left\{ {\begin{array}{*{20}{c}}
{\gamma _{{\rm{th,opt}}}^{{\rm{ip0}}} = \frac{{N\sigma _{0 }^2\sqrt {{n_a}\sigma _{0 }^4 + \left( {N - {n_a}} \right)\sigma _{1 }^4}  + \left( {{n_a}\sigma _{0 }^2 + \left( {N - {n_a}} \right)\sigma _{1 }^2} \right)\sqrt N \sigma _{0 }^2}}{{\sqrt N \sigma _{0 }^2 + \sqrt {{n_a}\sigma _{0 }^4 + \left( {N - {n_a}} \right)\sigma _{1 }^4} }},\left\{ {\begin{array}{*{20}{c}}
{{\rm{if}}\;\Delta n < 0{\rm{,}}B\left( {k - 1} \right) = 0}\\
{{\rm{if}}\;\Delta n > 0{\rm{,}}B\left( {k + 1} \right) = 0}
\end{array}} \right.}\\
{\gamma _{{\rm{th,opt}}}^{{\rm{ip1}}} = \frac{{\left( {{n_a}\sigma _{1 }^2 + \left( {N - {n_a}} \right)\sigma _{0 }^2} \right)\sqrt N \sigma _{1 }^2 + N\sigma _{1 }^2\sqrt {{n_a}\sigma _{1 }^4 + \left( {N - {n_a}} \right)\sigma _{0 }^4} }}{{\sqrt {{n_a}\sigma _{1 }^4 + \left( {N - {n_a}} \right)\sigma _{0 }^4}  + \sqrt N \sigma _{1 }^2}},\left\{ {\begin{array}{*{20}{c}}
{{\rm{if}}\;\Delta n < 0{\rm{,}}B\left( {k - 1} \right) = 1}\\
{{\rm{if}}\;\Delta n > 0{\rm{,}}B\left( {k + 1} \right) = 1}
\end{array}} \right.}
\end{array}} \right.,}\tag*{\textcolor{black}{(20)}}
\end{align}
\vspace{-15pt}
\hrulefill
\end{figure*}

{\textbf{Lemma 2.}} For ${n_a} \ge 0$, $P_{{\rm{BER}}}^{\rm{d}}$ strictly increases with the increasing of $n_a$.

{\emph{Remark 4.}} It is not hard to verify that $P_{{\rm{BER}}}^{\rm{d}}=0$ when $n_a=0$. Combining it with Lemma 2, it can be inferred that the RTSE increases the BER of the ED, and that $P_{{\rm{BER}}}^{\rm{d}}$ increases with $n_a$. This can be verified by Fig. \ref{fig4}. Particularly, as the SNR remains constant at 20 dB, and $n_a$ increases from 0 to 10, the BER increases from 0.00450097 to 0.0153945, indicating a serious performance degradation by $\frac{{0.0153945 - 0.00450097}}{{0.00450097}} \approx 2.42$ times. \textcolor{black}{The upper bound of the difference of BER can be obtained at ${n_a} = \frac{N}{2}$, which is given by
\vspace{-5pt}
\setcounter{equation}{14}
\begin{align}\label{15}
P_{{\rm{BER,upper}}}^{\rm{d}} = \frac{1}{4} - \frac{1}{2}Q\left( {\frac{{\sqrt N \left( {\sigma _{\max }^2 - \sigma _{\min }^2} \right)}}{{\sigma _{\min }^2 + \sigma _{\max }^2}}} \right).
\end{align}
\vspace{-5pt}
}

Thus, the RTSE should be considered when designing the symbol detector.
\section{Near-Optimal Symbol Detection Threshold under RTSE}
To alleviate the performance degradation caused by the RTSE, in this section, we first derive a closed-form expression for the near-optimal symbol detection threshold to minimize the BER under RTSE.
\subsection{Near-Optimal Symbol Detection Threshold}
In many previous works, e.g.,  \cite{7341107,8007328,7417704,7551180,10005249},
the ML criterion is used to obtain the optimal symbol detection threshold. However, such an approach may lead to  $\Pr \left( {\hat B\left( k \right) = 1|B\left( k \right) = 0} \right) \ne \Pr \left( {\hat B\left( k \right) = 0|B\left( k \right) = 1} \right)$, which is generally referred to as unbalanced BER\footnote{\textcolor{black}{The detailed explanation of how ML criterion causes an unbalanced BER under RTSE can be referred to Appendix E.}} \cite{6409345}. It may lead to BER deviations, which could affect reliability and stability of wireless communications particularly for transmission of large amounts of data. Therefore, we need to find an optimal detection threshold that can achieve balanced BER.

For the case with $B\left( {k - 1} \right) = 0$ and ${\Delta n < 0}$ (or $B\left( {k + 1} \right) = 0$ and ${\Delta n > 0}$), the balanced BER based optimal symbol detection threshold, denoted by ${\gamma _{{\rm{th,opt}}}^{{\rm{ip0}}}}$, is derived. Similarly, for the case with $B\left( {k - 1} \right) = 1$ and ${\Delta n < 0}$ (or $B\left( {k + 1} \right) = 1$ and ${\Delta n > 0}$), the balanced BER based optimal symbol detection threshold, denoted by ${\gamma _{{\rm{th,opt}}}^{{\rm{ip1}}}}$, is derived. {\color{black}{Both ${\gamma _{{\rm{th,opt}}}^{{\rm{ip0}}}}$ and ${\gamma _{{\rm{th,opt}}}^{{\rm{ip1}}}}$ can be obtained by solving \ref{16}, as shown at the top of this page, where ${{\tilde B}_{\Delta n}}\left( k \right) = B\left( {k + {\mathop{\rm sgn}} \left( {\Delta n} \right)} \right)$, ${\mathop{\rm sgn}} \left( {\Delta n} \right) = \left\{ {\begin{array}{*{20}{c}}
{1,{\rm{if}}\;\Delta n > 0}\\
{ - 1,{\rm{if}}\;\Delta n < 0}
\end{array}} \right.$, $\gamma _{{\rm{th,opt}}}^{{\rm{ip}}} = \left( {1 - i} \right)\gamma _{{\rm{th,opt}}}^{{\rm{ip0}}} + i\gamma _{{\rm{th,opt}}}^{{\rm{ip1}}}$, and $i \in \left\{ {0,1} \right\}$.

After mathematical transformations, \ref{16} becomes
\begin{align}\label{17}
\textcolor{black}{
\left\{ {\begin{array}{*{20}{c}}
{\int\limits_{\gamma _{{\rm{th,opt}}}^{{\rm{ip}}}}^\infty \!\! {{f_{\Gamma _{k|i,0}^{{\rm{ip}}}}}\left( z \right)dz}\!  = \!\!\!\!\int\limits_{ - \infty }^{\gamma _{{\rm{th,opt}}}^{{\rm{ip}}}}\!\! {{f_{\Gamma _{k|i,1}^{{\rm{ip}}}}}\left( z \right)dz} ,{\rm{if}}\;\sigma _0^2 \le \sigma _1^2}\\
{\int\limits_{ - \infty }^{\gamma _{{\rm{th,opt}}}^{{\rm{ip}}}} \!\! {{f_{\Gamma _{k|i,0}^{{\rm{ip}}}}}\left( z \right)dz} \! =\!\!\!\! \int\limits_{\gamma _{{\rm{th,opt}}}^{{\rm{ip}}}}^\infty \!\! {{f_{\Gamma _{k|i,1}^{{\rm{ip}}}}}\left( z \right)dz} ,{\rm{if}}\;\sigma _0^2 > \sigma _1^2}
\end{array}} \right..}\tag*{\textcolor{black}{(17)}}
\end{align}

According to \eqref{12} and \ref{17}, we can obtain the following two equations, given by
\setcounter{equation}{17}
\begin{align}\label{18}
Q\left( {\frac{{\gamma _{{\rm{th,opt}}}^{{\rm{ip0}}} \!\!-\!\! N\sigma _{0 }^2}}{{\sqrt N \sigma _{0 }^2}}} \right) \!\!=\!\! Q\left( {\frac{{\left( {{n_a}\sigma _{0 }^2 + \left( {N \!-\! {n_a}} \right)\sigma _{1 }^2} \right) \!\!-\!\! \gamma _{{\rm{th,opt}}}^{{\rm{ip0}}}}}{{\sqrt {{n_a}\sigma _{0 }^4 + \left( {N - {n_a}} \right)\sigma _{1 }^4} }}} \right),
\end{align}

\begin{align}\label{19}
Q\left( {\frac{{\gamma _{{\rm{th,opt}}}^{{\rm{ip1}}}\!\!-\!\! \left( {{n_a}\sigma _{1 }^2 + \left( {N \!-\! {n_a}} \right)\sigma _{0 }^2} \right)}}{{\sqrt {{n_a}\sigma _{1 }^4 + \left( {N - {n_a}} \right)\sigma _{0 }^4} }}} \right) \!\!=\!\! Q\left( {\frac{{N\sigma _{1 }^2 \!\!-\!\! \gamma _{{\rm{th,opt}}}^{{\rm{ip1}}}}}{{\sqrt N \sigma _{1 }^2}}} \right).
\end{align}

After mathematical transformations, the optimal symbol detection thresholds, $\gamma _{{\rm{th,opt}}}^{{\rm{ip0}}}$ and $\gamma _{{\rm{th,opt}}}^{{\rm{ip1}}}$, can be expressed as \ref{20}, as shown at the top of this page.}}

However, we note that the value of $B(k-1)$(or $B(k+1)$) cannot be obtained in practical AmBC. Due to the equiprobable symbols of the BT, we propose a weighted symbol detection threshold, which is referred to as the near-optimal symbol detection threshold in this paper, given by
\setcounter{equation}{20}
\begin{align}\label{21}\notag
\gamma _{{\rm{th,nopt}}}^{{\rm{ip}}} &= \frac{1}{2}\left( {\gamma _{{\rm{th,opt}}}^{{\rm{ip0}}} + \gamma _{{\rm{th,opt}}}^{{\rm{ip1}}}} \right) \\ \notag
&= \frac{1}{2}\frac{{N\sigma _{\min }^2\sqrt {{n_a}\sigma _{\min }^4 + \left( {N  - {n_a}} \right)\sigma _{\max }^4} }}{{\sqrt N \sigma _{\min }^2 + \sqrt {{n_a}\sigma _{\min }^4 + \left( {N - {n_a}} \right)\sigma _{\max }^4} }}\\ \notag
& + \frac{1}{2}\frac{{\left( {{n_a}\sigma _{\min }^2 + \left( {N - {n_a}} \right)\sigma _{\max }^2} \right)\sqrt N \sigma _{\min }^2}}{{\sqrt N \sigma _{\min }^2 + \sqrt {{n_a}\sigma _{\min }^4 + \left( {N - {n_a}} \right)\sigma _{\max }^4} }} \\ \notag
 &+ \frac{1}{2}\frac{{\left( {{n_a}\sigma _{\max }^2 + \left( {N - {n_a}} \right)\sigma _{\min }^2} \right)\sqrt N \sigma _{\max }^2}}{{\sqrt {{n_a}\sigma _{\max }^4 + \left( {N - {n_a}} \right)\sigma _{\min }^4}  + \sqrt N \sigma _{\max }^2}}\\ \notag
 &+ \frac{1}{2}\frac{{N\sigma _{\max }^2\sqrt {{n_a}\sigma _{\max }^4 + \left( {N - {n_a}} \right)\sigma _{\min }^4} }}{{\sqrt {{n_a}\sigma _{\max }^4 + \left( {N - {n_a}} \right)\sigma _{\min }^4}  + \sqrt N \sigma _{\max }^2}}. \tag{21}
\end{align}

{\emph{Remark 5.}} The accuracy of the near-optimal symbol detection threshold should be ensured, which will be verified in Fig. \ref{fig5} and Table \ref{table2}. We note that the near-optimal symbol detection threshold avoids knowing $B(k-1)$ and $B(k+1)$, yet the parameters ${\sigma _{\min}^2}$ and ${\sigma _{\max}^2}$ and the RTSE $n_a$, which are unknown to the BR in practice, are still required. To address it, we propose a novel method to estimate these two parameters and the RTSE $n_a$ using the received signal samples at the BR in the next subsection.

\subsection{Symbol Detection Threshold Estimation}
Here we estimate ${\sigma _{\min}^2}$, ${\sigma _{\max}^2}$ and $n_a$ by  exploiting the attributes of the received samples\footnote{Although jointly leveraging these samples and pilot symbols can enhance the estimation accuracy of the symbol detection threshold, addressing this approach poses a new challenge that we leave for future work. } that are obtained after TS within one transmission block, and the corresponding method is summarized in Algorithm 1. Such a method has been widely used in the existing works such as \cite{7551180} and \cite{10005249}.
\begin{table}[!ht]
\centering
   \begin{tabular}{l}
      \toprule
      \textbf{Algorithm 1} The parameter estimation method under RTSE \\
      \midrule
      \textbf{Input:} Received samples $y_k^{\rm{ip}}(n)$, $N$ and $K$\\
      \textbf{Output:} The parameter $\hat n_a$, ${\hat \sigma _{\min }^2}$ and ${\hat \sigma _{\max }^2}$\\
      \: 1: \textbf{Step 1: Power Calculation}\\
      \: 2: Compute power of samples corresponding to $B(k)$:\\
      \: 3: ${A_k} = \frac{1}{N}\sum\limits_{n = \left( {k - 1} \right)N + 1}^{kN} {{{\left| {y_k^{{\rm{ip}}}\left( n \right)} \right|}^2}}$\\
      \: 4: \textbf{Step 2: Statistical Analysis}\\
      \: 5: Sort $\left\{ {{A_k}} \right\}_{k = 1}^K$ in ascending order to obtain $\left\{ {A_k^ \uparrow } \right\}_{k = 1}^K$ \\
      \: 6: Divide $\left\{ {A_k^ \uparrow } \right\}_{k = 1}^K$ into four equal parts: \\
      \: 7: $\left\{ {A_k^ \uparrow } \right\}_{k = 1}^{{K \mathord{\left/
 {\vphantom {K 4}} \right.
 \kern-\nulldelimiterspace} 4}}$, $\left\{ {A_k^ \uparrow } \right\}_{k \!=\! {K \mathord{\left/
 {\vphantom {K 4}} \right.
 \kern-\nulldelimiterspace} 4} \!+\! 1}^{{K \mathord{\left/
 {\vphantom {K 2}} \right.
 \kern-\nulldelimiterspace} 2}}$, $\left\{ {A_k^ \uparrow } \right\}_{k \!=\! {K \mathord{\left/
 {\vphantom {K 2}} \right.
 \kern-\nulldelimiterspace} 2} \!+\! 1}^{{{3K} \mathord{\left/
 {\vphantom {{3K} 4}} \right.
 \kern-\nulldelimiterspace} 4}}$ and $\left\{ {A_k^ \uparrow } \right\}_{k \!=\! {{3K} \mathord{\left/
 {\vphantom {{3K} 4}} \right.
 \kern-\nulldelimiterspace} 4} \!+\! 1}^K$ \\
     \: 8: Compute average power of each part:\\
     \: 9: ${E_1}$, ${E_2}$, ${E_3}$ and ${E_4}$\\
     \: 10: \textbf{Step 3: Parameter Estimation}\\
     \: 11: Estimate ${\sigma _{\min }^2}$ and ${\sigma _{\max }^2}$: \\
     \: 12: $\hat \sigma _{\min }^2 \!\!=\!\! {E_1} \!\!=\!\! \frac{4}{K}\sum\limits_{k = 1}^{{K \mathord{\left/
 {\vphantom {K 4}} \right.
 \kern-\nulldelimiterspace} 4}} {A_k^ \uparrow },
\hat \sigma _{\max }^2 \!\!=\!\! {E_4} \!\!=\!\! \frac{4}{K}\sum\limits_{k = {{3K} \mathord{\left/
 {\vphantom {{3K} 4}} \right.
 \kern-\nulldelimiterspace} 4} + 1}^K {A_k^ \uparrow }$\\
     \: 13: Estimate $n_a$:\\
     \: 14: ${\hat n_a} = \frac{N}{2}\left( {1 - \frac{{{E_3} - {E_2}}}{{\hat \sigma _{\max }^2 - \hat \sigma _{\min }^2}}} \right)$\\
     %\: 15: Transmit pilot symbols to compare sizes of $\sigma _0^2$ and $\sigma _1^2$, as in \cite{8007328}\\
     \: 15: \textbf{return} $\hat n_a$, ${\hat \sigma _{\min }^2}$ and ${\hat \sigma _{\max }^2}$ \\
      \bottomrule
   \end{tabular}
\end{table}

\begin{figure*}[t]
\normalsize
\begin{align}\label{28}\notag
\textcolor{black}{
\hat \gamma _{{\rm{th,nopt}}}^{{\rm{ip,PSK}}}} &\textcolor{black}{= \frac{1}{2}\frac{{N\hat \sigma _{\min }^2\sqrt {{{\hat n}_a}\left( {2\hat \sigma _{\min }^2 - {N_\omega }} \right) + \left( {N - {{\hat n}_a}} \right)\left( {2\hat \sigma _{\max }^2 - {N_\omega }} \right)}  + \left( {{{\hat n}_a}\hat \sigma _{\min }^2 + \left( {N - {{\hat n}_a}} \right)\hat \sigma _{\max }^2} \right)\sqrt {N\left( {2\hat \sigma _{\min }^2 - {N_\omega }} \right)} }}{{\sqrt {N\left( {2\hat \sigma _{\min }^2 - {N_\omega }} \right)}  + \sqrt {{{\hat n}_a}\left( {2\hat \sigma _{\min }^2 - {N_\omega }} \right) + \left( {N - {{\hat n}_a}} \right)\left( {2\hat \sigma _{\max }^2 - {N_\omega }} \right)} }}} \\ \notag
 & \textcolor{black}{  + \frac{1}{2}\frac{{\left( {{{\hat n}_a}\hat \sigma _{\max }^2 + \left( {N - {{\hat n}_a}} \right)\hat \sigma _{\min }^2} \right)\sqrt {N\left( {2\hat \sigma _{\max }^2 - {N_\omega }} \right)}  + N\hat \sigma _{\max }^2\sqrt {{{\hat n}_a}\left( {2\hat \sigma _{\max }^2 - {N_\omega }} \right) + \left( {N - {{\hat n}_a}} \right)\left( {2\hat \sigma _{\min }^2 - {N_\omega }} \right)} }}{{\sqrt {{{\hat n}_a}\left( {2\hat \sigma _{\max }^2 - {N_\omega }} \right) + \left( {N - {{\hat n}_a}} \right)\left( {2\hat \sigma _{\min }^2 - {N_\omega }} \right)}  + \sqrt {N\left( {2\hat \sigma _{\max }^2 - {N_\omega }} \right)} }},}  \tag*{\textcolor{black}{(28)}}
\end{align}
\vspace{-5pt}
\hrulefill
\end{figure*}

\setcounter{table}{1}
\begin{table*}[ht]
    \centering
    \caption{Four cases under RTSE.}
    \label{table1}
    \begin{tabular}{|c|c|c|c|c|}
        \hline
        $B(k-1)$ & $B(k)$ & Power & $\sigma _0^2 \le \sigma _1^2$ & $\sigma _0^2 > \sigma _1^2$ \\
        \hline
        0 & 0 & ${E_{00}} = \sigma _0^2$ & \multirow{2}{*}{${E_{00}} < {E_{10}} < {E_{01}} < {E_{11}}$} & \multirow{2}{*}{${E_{11}} < {E_{01}} < {E_{10}} < {E_{00}}$} \\
        \cline{1-3}
        1 & 0 & ${E_{10}} = \frac{{{n_a}\sigma _1^2 + \left( {N - {n_a}} \right)\sigma _0^2}}{N}$ & \multirow{2}{*}{${E_1} \simeq {E_{00}} = {\sigma _0^2},{E_2} \simeq {E_{10}}$} & \multirow{2}{*}{${E_1} \simeq {E_{11}} = {\sigma _1^2},{E_2} \simeq {E_{01}}$}\\
        \cline{1-3}
        0 & 1 & ${E_{01}} = \frac{{{n_a}\sigma _0^2 + \left( {N - {n_a}} \right)\sigma _1^2}}{N}$ & \multirow{2}{*}{${E_3} \simeq {E_{01}},{E_4} \simeq {E_{11}} = \sigma _1^2$} & \multirow{2}{*}{${E_3} \simeq {E_{10}},{E_4} \simeq {E_{00}} = \sigma _0^2$}\\
        \cline{1-3}
        1 & 1 & ${E_{11}} = \sigma _1^2$ & \multirow{2}{*}{} & \multirow{2}{*}{}\\
        \hline
    \end{tabular}
    \vspace{-15pt}
\end{table*}

\emph{Proof of line 12 in Algorithm 1.} When $\Delta n < 0$ and using the equal probability transmission of `0' and `1', we can identify four cases based on the BT's symbols, as shown in Table \ref{table1}. Clearly, the average power of the first part equals ${\hat \sigma _{\min }^2}$, and the average power of the fourth part equals ${\hat \sigma _{\max }^2}$. The same conclusion can be obtained when $\Delta n > 0$.

\emph{Proof of line 14 in Algorithm 1.} Please refer to Appendix F.  \hfill {$\blacksquare $}

Once $\hat n_a$, ${\hat \sigma _{\min }^2}$, and ${\hat \sigma _{\max }^2}$ are obtained, the near-optimal symbol detection threshold  can be estimated as
\begin{align}\label{22}\notag
\hat \gamma _{{\rm{th,nopt}}}^{{\rm{ip}}} &= \frac{1}{2}\frac{{N\hat\sigma _{ \min }^2\sqrt {{\hat n_a}\hat \sigma _{\min }^4 + \left( {N  - {\hat n_a}} \right)\hat \sigma _{\max }^4} }}{{\sqrt N \hat \sigma _{\min }^2 + \sqrt {{\hat n_a}\hat \sigma _{\min }^4 + \left( {N - {\hat n_a}} \right)\hat \sigma _{\max }^4} }}\\ \notag
& + \frac{1}{2}\frac{{\left( {{\hat n_a}\hat \sigma _{\min }^2 + \left( {N - {\hat n_a}} \right)\hat \sigma _{\max }^2} \right)\sqrt N \hat \sigma _{\min }^2}}{{\sqrt N \hat \sigma _{\min }^2 + \sqrt {{\hat n_a}\hat \sigma _{\min }^4 + \left( {N - {\hat n_a}} \right)\hat \sigma _{\max }^4} }}\\ \notag
 &+ \frac{1}{2}\frac{{\left( {{\hat n_a}\hat \sigma _{\max }^2 + \left( {N - {\hat n_a}} \right)\hat \sigma _{\min }^2} \right)\sqrt N \hat \sigma _{\max }^2}}{{\sqrt {{\hat n_a}\hat \sigma _{\max }^4 + \left( {N - {\hat n_a}} \right)\hat \sigma _{\min }^4}  + \sqrt N\hat \sigma _{\max }^2}}\\
 &+ \frac{1}{2}\frac{{N\hat \sigma _{\max }^2\sqrt {{\hat n_a}\hat\sigma _{\max }^4 + \left( {N - {\hat n_a}} \right)\hat \sigma _{\min }^4} }}{{\sqrt {{\hat  n_a}\hat \sigma _{\max }^4 + \left( {N - {\hat n_a}} \right)\hat \sigma _{\min }^4}  + \sqrt N \hat \sigma _{\max }^2}}. \tag{22}
\end{align}

\textcolor{black}{\emph{Remark 6.} According to \eqref{8}, \eqref{9}, and \eqref{22}, the computational complexity of the ED under \textcolor{black}{RTSE} is $O\left( {KN + K\log K} \right)$.}

\color{black}
\subsection{The ED with a PSK Signal}
In practice, S may emit a PSK signal rather than a complex Gaussian signal. In this section, we analyze the ED performance with a PSK signal under RTSE.

The PSK signal can be expressed as
\setcounter{equation}{22}
\begin{align}\label{23}
s\left( n \right) = \sqrt {{P_s}} \exp \left( {j{{2\pi k} \mathord{\left/
 {\vphantom {{2\pi k} M}} \right.
 \kern-\nulldelimiterspace} M}} \right),k = 0, \cdots ,M - 1,
\end{align}%s\left( n \right) = \sqrt {{P_s}} \exp \left( {j\frac{{2\pi k}}{M}} \right)
where $P_s$ is the signal power.

The BER and the optimal symbol detection threshold for PSK signal under perfect TS can be, respectively, derived as
\begin{align}\label{24}
P_{{\rm{BER}}}^{{\rm{p,PSK}}} \!\!=\!\! \frac{1}{2}Q\left( {\frac{{\gamma _{{\rm{th}}}^{{\rm{p,PSK}}} \!\!-\!\! N\sigma _{\min }^2}}{{\sqrt {N{\xi _{\min }}} }}} \right) \!\!+\!\! \frac{1}{2}Q\left( {\frac{{N\sigma _{\max }^2 \!\!-\!\! \gamma _{{\rm{th}}}^{{\rm{p,PSK}}}}}{{\sqrt {N{\xi _{\max }}} }}} \right),
\end{align}

\begin{align}\label{25}
\gamma _{{\rm{th,opt}}}^{{\rm{p,PSK}}} = \frac{{N\left( {\sigma _{\min }^2\sqrt {{\xi _{\max }}}  + \sigma _{\max }^2\sqrt {{\xi _{\min }}} } \right)}}{{\sqrt {{\xi _{\min }}}  + \sqrt {{\xi _{\max }}} }},
\end{align}
where ${\gamma _{{\rm{th}}}^{{\rm{p,PSK}}}}$ is the symbol
detection threshold under perfect TS, ${\xi _0} = {N_\omega }\left( {2\sigma _0^2 - {N_\omega }} \right)$, ${\xi _1} = {N_\omega }\left( {2\sigma _1^2 - {N_\omega }} \right)$, ${\xi _{\min }} = \min \left\{ {{\xi _0},{\xi _1}} \right\}$ and ${\xi _{\max }} = \max \left\{ {{\xi _0},{\xi _1}} \right\}$. When the optimal symbol detection threshold \eqref{25} is used, the balanced BER can be achieved.

Similar to the analysis of the complex Gaussian signal, the BER and the near-optimal symbol detection threshold for PSK signal under \textcolor{black}{RTSE} can be, respectively, derived as
\begin{align}\label{26}\notag
P_{{\rm{BER}}}^{{\rm{ip,PSK}}} &\simeq \frac{1}{4}Q\left( {\frac{{\gamma _{{\rm{th}}}^{{\rm{ip,PSK}}} \!\!\!\!-\!\! N\sigma _{\min }^2}}{{\sqrt {N{\xi _{\min }}} }}} \right) \!\!+\!\! \frac{1}{4}Q\left( {\frac{{N\sigma _{\max }^2 \!\!\!\!-\!\! \gamma _{{\rm{th}}}^{{\rm{ip,PSK}}}}}{{\sqrt {N{\xi _{\max }}} }}} \right)\notag \\
& + \frac{1}{4}Q\left( {\frac{{\gamma _{{\rm{th}}}^{{\rm{ip,PSK}}} \!\!-\!\! \left( {{n_a}\sigma _{\max }^2 \!\!+\!\! \left( {N \!\!-\!\! {n_a}} \right)\sigma _{\min }^2} \right)}}{{\sqrt {{n_a}{\xi _{\max }} + \left( {N - {n_a}} \right){\xi _{\min }}} }}} \right)\notag \\
& + \frac{1}{4}Q\left( {\frac{{\left( {{n_a}\sigma _{\min }^2 \!\!+\!\! \left( {N \!\!-\!\! {n_a}} \right)\sigma _{\max }^2} \right) \!\!-\!\! \gamma _{{\rm{th}}}^{{\rm{ip,PSK}}}}}{{\sqrt {{n_a}{\xi _{\min }} + \left( {N - {n_a}} \right){\xi _{\max }}} }}} \right),\tag{26}
\end{align}

\begin{align}\label{27}\notag
\gamma _{{\rm{th,nopt}}}^{{\rm{ip,PSK}}} &= \frac{1}{2}\frac{{N\sigma _{\min }^2\sqrt {{n_a}{\xi _{\min }} + \left( {N - {n_a}} \right){\xi _{\max }}} }}{{\sqrt {N{\xi _{\min }}}  + \sqrt {{n_a}{\xi _{\min }} + \left( {N - {n_a}} \right){\xi _{\max }}} }} \\ \notag
& + \frac{1}{2}\frac{{\left( {{n_a}\sigma _{\min }^2 + \left( {N - {n_a}} \right)\sigma _{\max }^2} \right)\sqrt {N{\xi _{\min }}} }}{{\sqrt {N{\xi _{\min }}}  + \sqrt {{n_a}{\xi _{\min }} + \left( {N - {n_a}} \right){\xi _{\max }}} }}\\ \notag
& + \frac{1}{2}\frac{{\left( {{n_a}\sigma _{\max }^2 + \left( {N - {n_a}} \right)\sigma _{\min }^2} \right)\sqrt {N{\xi _{\max }}} }}{{\sqrt {{n_a}{\xi _{\max }} + \left( {N - {n_a}} \right){\xi _{\min }}}  + \sqrt {N{\xi _{\max }}} }}\\ \notag
 &  + \frac{1}{2}\frac{{N\sigma _{\max }^2\sqrt {{n_a}{\xi _{\max }} + \left( {N - {n_a}} \right){\xi _{\min }}} }}{{\sqrt {{n_a}{\xi _{\max }} + \left( {N - {n_a}} \right){\xi _{\min }}}  + \sqrt {N{\xi _{\max }}} }}, \tag{27}
\end{align}
where ${\gamma _{{\rm{th}}}^{{\rm{ip,PSK}}}}$ is the symbol
detection threshold for PSK signal under \textcolor{black}{RTSE}.

According to the parameter estimation method proposed in Section IV-B, once $\hat n_a$, ${\hat \sigma _{\min }^2}$, and ${\hat \sigma _{\max }^2}$ are obtained, the near-optimal symbol detection threshold for PSK signal can be estimated as \ref{28}, as shown at the top of this page.
\color{black}
\section{Simulation Results}
\setcounter{table}{0}
\begin{table*}[t]
    \centering
    \caption{Thresholds comparison of different RTSE.}
    \label{table2}
    \begin{tabular}{|c|c|c|c|c|}
        \hline
        RTSE & Our theoretical $\gamma _{{\rm{th,nopt}}}^{{\rm{ip}}}$ & Our estimated $\hat \gamma _{{\rm{th,nopt}}}^{{\rm{ip}}}$ & Theoretical $\gamma _{{\rm{th,opt}}}^{{\rm{p}}}$ in \cite{8007328} & Estimated $\gamma _{{\rm{th,opt}}}^{{\rm{p}}}$ in \cite{8007328}\\
        \hline
        $n_a=0$ & 12558 & 12871 & 12558 & 12562 \\
        \hline
        $n_a=10$ & 12743 & 12992 & Not available & Not available \\
        \hline
        $n_a=20$ & 12901 & 13083 & Not available & Not available \\
        \hline
    \end{tabular}
    \vspace{-10pt}
\end{table*}

In this section, computer simulations are presented to support our analytical results. All channel coefficients $h$, $f$ and $g$ are defined as \textcolor{black}{$h,f,g \sim \mathbb{CN}\left( {0,1} \right)$}. The complex signal attenuation within the BT is consistently set as 1.1 dB \cite{kellogg2016passive}. The noise variance ${N_\omega } $ and the number of BT's symbol in one transmission block $K$ are set as 1 and 100, respectively.
\subsection{Verification of Derived Results and Remark 4}
In this subsection, we verify the correctness of our derived PDF under RTSE.

Fig. \ref{fig3} plots the empirical PDF of $\Gamma _k^{\rm{ip}}$ and the analytical PDF of $\Gamma _k^{\rm{ip}}$ under RTSE. We set SNR = 20 dB, $N=100$, and  ${n_a} = 10$. One can see that the analytical PDF of $\Gamma _k^{\rm{ip}}$, as derived in \eqref{12}, closely aligns with the empirical PDF of $\Gamma _k^{\rm{ip}}$ under RTSE. This is true regardless of whether ${\sigma _0^2 \le \sigma _1^2}$ or ${\sigma _0^2 > \sigma _1^2}$. Additionally, a serious deviation can be observed between ${f_{\Gamma _{k|0,0}^{{\rm{ip}}}}}\left( z \right)$ and ${f_{\Gamma _{k|1,0}^{{\rm{ip}}}}}\left( z \right)$, ${f_{\Gamma _{k|1,1}^{{\rm{ip}}}}}\left( z \right)$ and ${f_{\Gamma _{k|0,1}^{{\rm{ip}}}}}\left( z \right)$, demonstrating that the PDF under RTSE is different from that under perfect TS. This observation validates the accuracy of our derived PDF of $\Gamma _k^{\rm{ip}}$ given by \eqref{12} under RTSE.

\setcounter{figure}{4}
\begin{figure*}
			\centering   %居中放置
			\subfigure[${\sigma _0^2 \le \sigma _1^2}$.] % 为每个图片加上编号
			{
				\begin{minipage}[b]{0.38\linewidth} % 表示在该行所占空间的比例为0.3；其中。3=0.3；b为放置方式；
					\centering
					\includegraphics[width=1\textwidth]{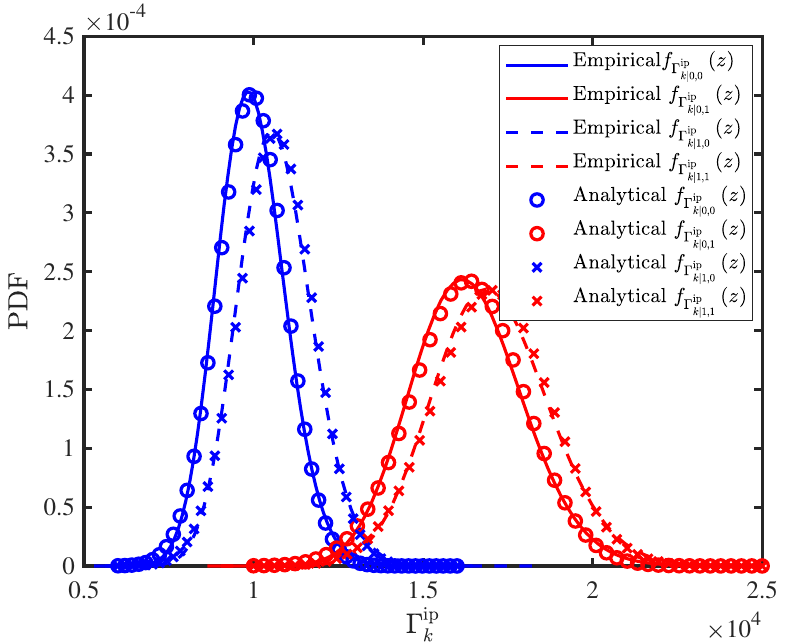}
                \label{fig3_a}
				\end{minipage}
			}
			\subfigure[${\sigma _0^2 > \sigma _1^2}$.]
			{
				\begin{minipage}[b]{0.38\linewidth}
					\centering
					\includegraphics[width=1\textwidth]{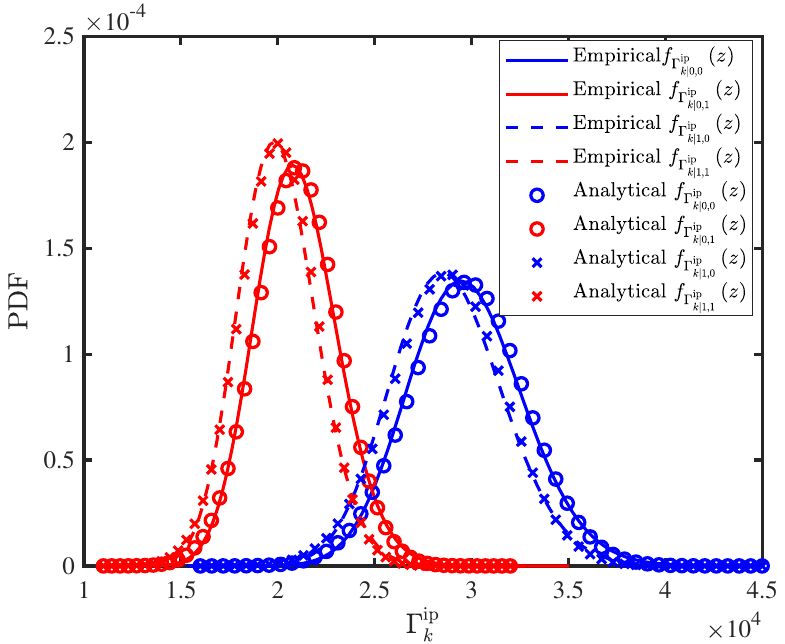}
                \label{fig3_b}
				\end{minipage}
			}
			\caption{Empirical and analytical PDFs for $\Gamma _k^{{\rm{ip}}}$.}
        \label{fig3}
        \vspace{-10pt}
\end{figure*}

To demonstrate the accuracy of \eqref{13} and illustrate \emph{Remark 4}, we plot Fig. \ref{fig4}, where `with perfect TS (`p')' and `with RTSE (`ip')' represent the cases under perfect TS and RTSE, respectively. We \textcolor{black}{set $N=100$ and} adopt the symbol detection threshold \eqref{5} for all cases. Theoretical BER in \eqref{13} is also provided for comparison with simulation results. It can be seen that the simulation results with `ip' always match well with the theoretical results in \eqref{13}, which indicates the correctness of the theoretical derivations. It can also be seen that the BER with `ip' is always higher than that with `p', and the BER increases as $n_a$ increases, which indicates the RTSE degrades the detection performance of ED. This is because the change in the symbol detection model due to RTSE has led to a change in the PDF, which further leads to a change in the symbol detection threshold. In such a case, if we continue to use the symbol detection threshold provided by \eqref{5} under RTSE, the BER will greatly increase, as expected in \emph{Remark 4}. %Furthermore, higher SNR results in lower BER, and the BER flattens out as the SNR becomes relatively large for all cases. The reason for that lies in the fact that the BR suffers from the direct link interference caused by the ambient source.
\setcounter{figure}{2}
\begin{figure}[h]
  \centering
  \includegraphics[width=0.38\textwidth]{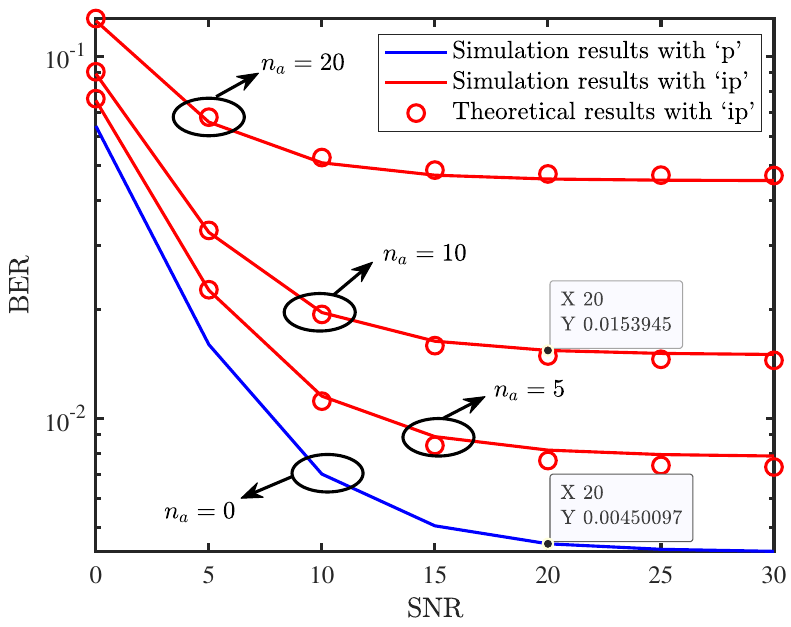}\\
  \caption{\textcolor{black}{BER versus SNR with the  symbol detection threshold \eqref{5}.}}\label{fig4}
  \vspace{-15pt}
\end{figure}
\subsection{Verification of the Obtained Symbol Detection Threshold}
\begin{figure}
  \centering
  \includegraphics[width=0.38\textwidth]{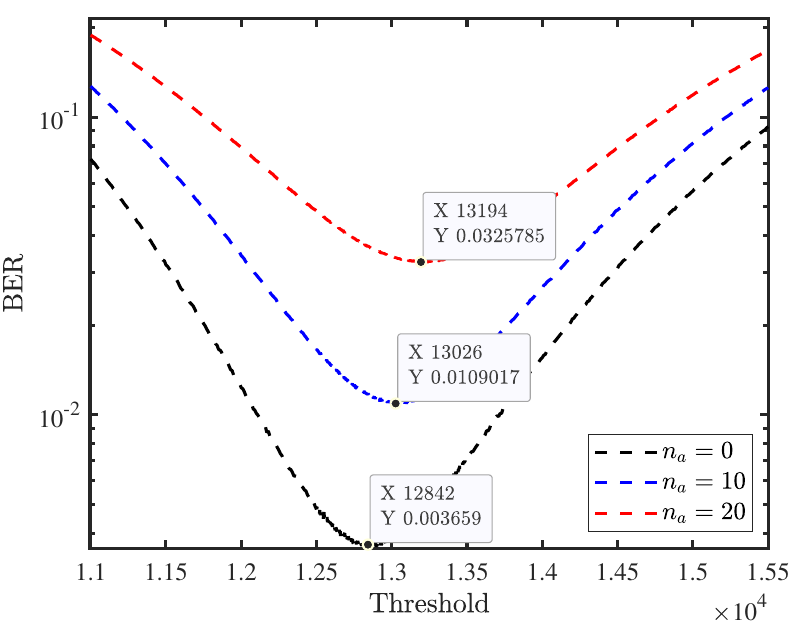}\\
  \caption{Verification of the near-optimality of \eqref{21} and  \eqref{22}.}\label{fig5}
  \vspace{-15pt}
\end{figure}

To verify the near-optimality of the symbol detection thresholds \eqref{21} and \eqref{22}, the relationships between BER and symbol detection threshold under various RTSE are given in Fig. \ref{fig5}. We set $N = 100$ and SNR = 20 dB. It is clear  that there is an optimal symbol detection threshold minimizing BER. Table \ref{table2} presents theoretical and estimated near-optimal symbol detection thresholds, calculated by \eqref{21} and \eqref{22}, respectively. In contrast, those from \cite{8007328} are calculated using \eqref{5} with their exact and estimated values of ${\sigma _0^2}$ and ${\sigma _1^2}$, respectively. Comparing optimal symbol detection threshold under $n_a=0$  in Fig. \ref{fig5} with that listed in Table \ref{table2}, it becomes apparent that our theoretical and estimated symbol detection thresholds more closely approximate the optimal one compared to those in \cite{8007328}.

The accuracy of the estimated detection threshold $\hat \gamma _{{\rm{th,nopt}}}^{{\rm{ip}}}$ directly affects the symbol detection performance. Therefore, we define the accuracy of the estimated detection threshold as $\gamma _{{\rm{th,nopt}}}^{{\rm{diff}}} = \frac{{\left| {\hat \gamma _{{\rm{th,nopt}}}^{{\rm{ip}}} - \gamma _{{\rm{th,nopt}}}^{{\rm{ip}}}} \right|}}{{\gamma _{{\rm{th,nopt}}}^{{\rm{ip}}}}}$. Fig. \ref{fig55} shows the effect of $N$ and $K$ on $\gamma _{{\rm{th,nopt}}}^{{\rm{diff}}}$, where we set SNR = 20 dB and $n_a = 20$. It is observed that the accuracy of our estimated detection threshold is  high even with a small number of samples (i.e., $N=50$). For a given $N$, it can also be seen that as the symbol number $ K $ increases, the estimated $\hat \gamma _{{\rm{th,nopt}}}^{{\rm{ip}}}$ becomes more accurate. Furthermore, as $N$ increases, the accuracy of the estimated detection threshold also increases.

\setcounter{figure}{5}
\begin{figure}[h]
  \centering
  \includegraphics[width=0.38\textwidth]{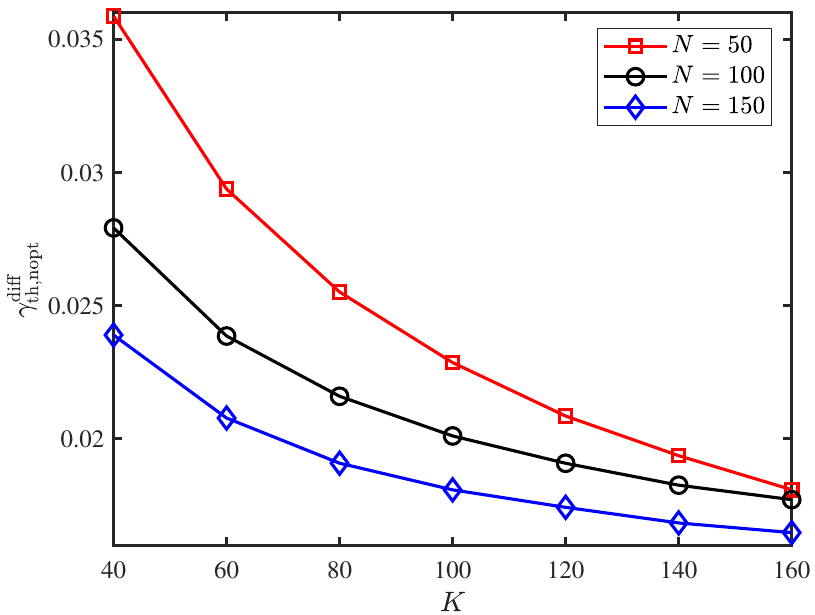}\\
  \centering \caption{$\gamma _{{\rm{th,nopt}}}^{{\rm{diff}}}$ versus $K$ under different $N$.}\label{fig55}
  \vspace{-10pt}
\end{figure}

\begin{figure}[h]
  \centering
  \includegraphics[width=0.38\textwidth]{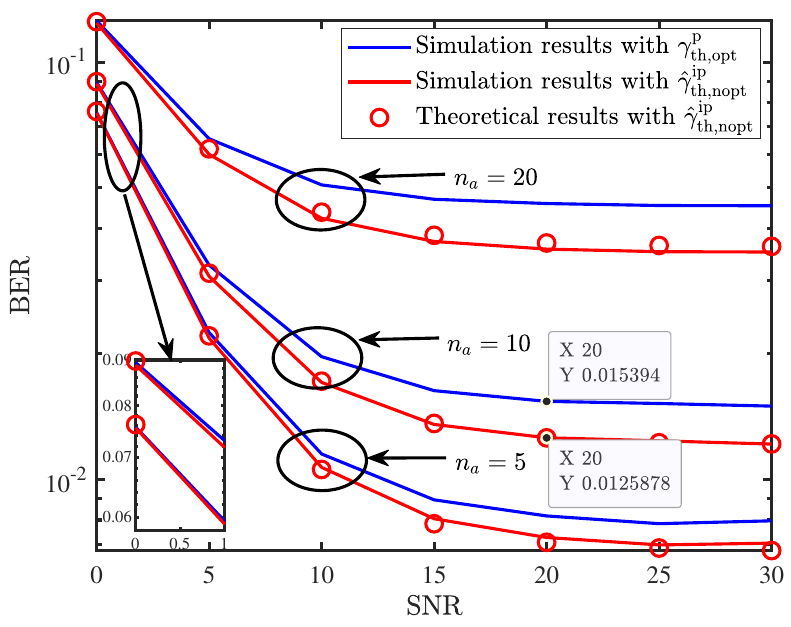}\\
  \caption{BER comparison between  ED with  the symbol detection threshold $\gamma _{{\rm{th,opt}}}^{{\rm{p}}}$ in \eqref{5} and that with the symbol detection threshold  $\hat \gamma _{{\rm{th,nopt}}}^{{\rm{ip}}}$ in \eqref{22}.}\label{fig6}
  \vspace{-10pt}
\end{figure}

\begin{figure}[h]
  \centering
  \includegraphics[width=0.365\textwidth]{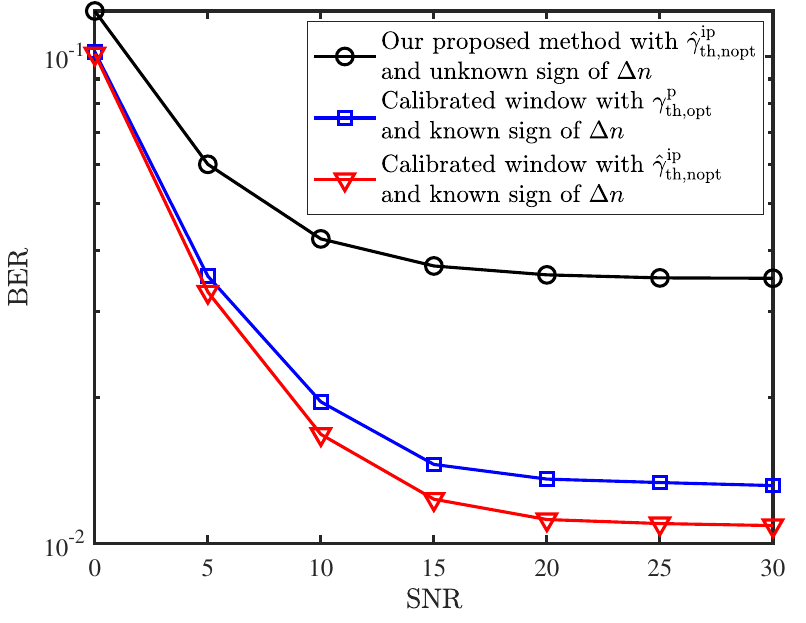}\\
  \caption{{The comparison between  our proposed method and ED.}}\label{fig0}
  \vspace{-5pt}
\end{figure}

\begin{figure}[h]
  \centering
  \includegraphics[width=0.38\textwidth]{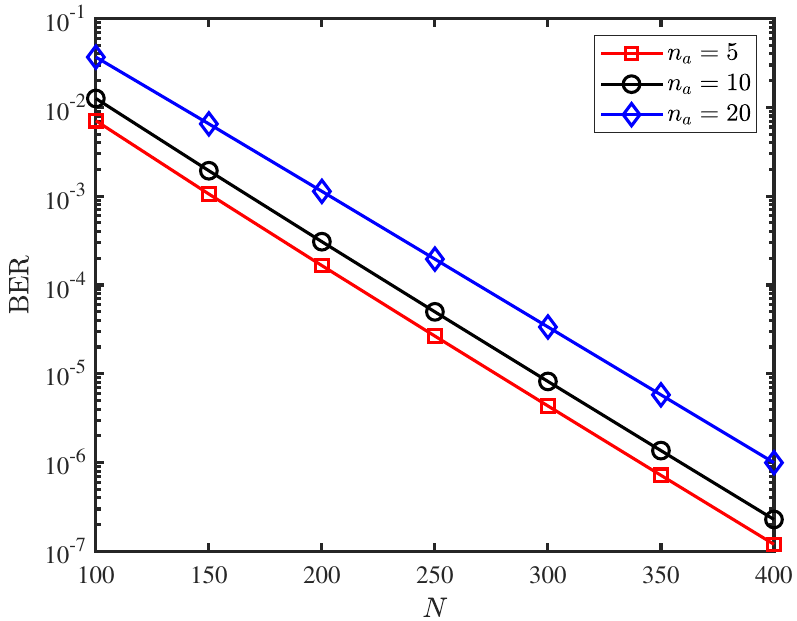}\\
  \centering \caption{BER versus $N$ with the  near-optimal symbol detection threshold.}\label{fig7}
  \vspace{-10pt}
\end{figure}

\begin{figure}[h]
  \centering
  \includegraphics[width=0.38\textwidth]{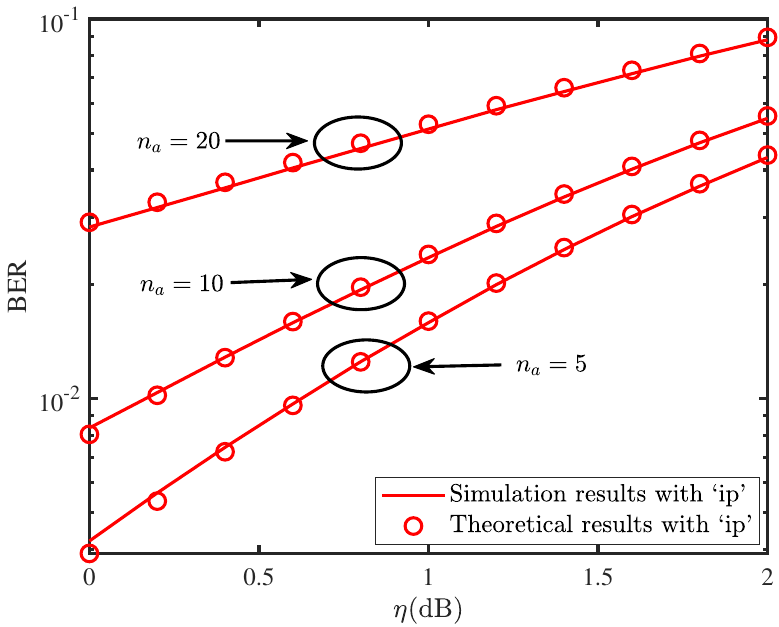}\\
  \centering \caption{BER versus $\eta$ with the near-optimal symbol detection threshold.}\label{fig8}
  \vspace{-10pt}
\end{figure}

Fig. \ref{fig6} plots the BER versus SNR. We set $N = 100$ for each case. We use our derived near-optimal symbol detection threshold $\hat \gamma _{{\rm{th,nopt}}}^{{\rm{ip}}}$, shown in  \eqref{22}, and the optimal symbol detection threshold $\gamma _{{\rm{th,opt}}}^{{\rm{p}}}$ \cite{8007328}, shown in \eqref{5} in this paper, under RTSE, respectively. It can be observed that our derived near-optimal symbol detection threshold $\hat \gamma _{{\rm{th,nopt}}}^{{\rm{ip}}}$ achieves a smaller BER compared to the optimal symbol detection threshold $\gamma _{{\rm{th,opt}}}^{{\rm{p}}}$ in \cite{8007328}. This proves that our derived symbol detection threshold improves BER performance of ED. Specifically, with SNR = 20 dB and $n_a=10$, the BER decreases from 0.015394 to 0.0125878, indicating a significant performance improvement of approximately $18\%$: $\frac{{0.015394 - 0.0125878}}{{0.015394}} \times 100\%  \approx 18\% $. It can also be observed that as $n_a$ increase, the difference of BER between RTSE and perfect TS exhibits a corresponding rise, indicating the effectiveness of $\hat \gamma _{{\rm{th,nopt}}}^{{\rm{ip}}}$ in reducing BER becomes more pronounced. However, similar to the perfect TS case in \cite{8007328}, an BER floor exists due to the RTSE $n_a$ and the difference in channel coefficients ${\left( {{{\left| \mu  \right|}^2} - {{\left| h \right|}^2}} \right)}$\footnote{The detailed reasons for the existence of the BER floor can be referred to Appendix G.}.

Fig. \ref{fig0} compares the detection performance of our proposed method  with the ED. For our proposed method, we consider two cases, i.e.,  the  sign of  $\Delta n$ is known or unknown at the BR, while for the ED, we consider the BR knows the sign of $\Delta n$. Please note that if the sign of $\Delta n$ is known, the BR can reduce the RTSE via calibrated window method before symbol detection, and thus improving the detection performance.  It can be seen that the knowledge of the sign brings significant performance gains for our proposed method. Notably,  with the sign known, our method consistently outperforms the calibrated-window ED. This is because the calibrated window can only mitigate, but not eliminate, the RTSE. In contrast, our symbol detection threshold is explicitly derived to account for the residual RTSE, whereas the ED's threshold is not, leading to its higher BER.

\color{black}
Fig. \ref{fig7} shows the effect of $N$ on BER, where we set SNR = 20 dB, {\color{black}$N$ is range\footnote{ $N>100$ is practical in AmBC due to the large rate difference between BT and ambient RF source. Particularly, according to Release 18 of 3GPP \cite{10463656}, the supported data rates of AmBC are in the range of 0.1kbps-5kbps, whereas some ambient RF sources, such as cellular network base stations, can achieve data rates in the order of several Mbps. Consequently, the ratio between the cellular user's data rate and the BT's rate can be in the thousands, suggesting that $N$ can be in the thousands or even larger than 1000 in practical AmBC. }
 from 100 to 400}, and use \eqref{21} to calculate $\gamma_{{\rm{th,nopt}}}^{{\rm{ip}}}$. It can be seen that as $N$ increases, the BER decreases, indicating that increasing the number of samples can effectively reduce BER.

Fig. \ref{fig8} illustrates the impact of signal's complex attenuation inside the BT $\eta$ on BER performance under a fixed SNR of 15 dB with $N=100$. In the simulation, $\eta$ is set to range from 0 dB to 2 dB. Both theoretical and simulated BER results show excellent consistency across the entire range of $\eta$. Notably, the BER increases as $\eta$ increases for all $n_a$, indicating that $\eta$ plays a significant role in influencing system performance.

\subsection{Verification of ED with a PSK Signal}
\begin{figure}
  \centering
  \includegraphics[width=0.38\textwidth]{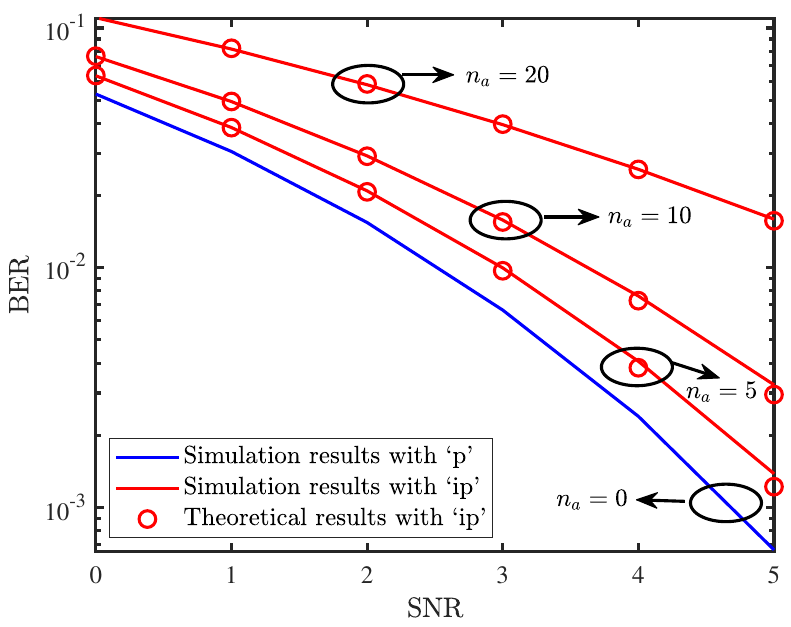}\\
  \centering \caption{\textcolor{black}{BER versus SNR with the symbol detection threshold $\gamma _{{\rm{th,opt}}}^{{\rm{p,PSK}}}$ \eqref{25} when PSK signal is used.}}\label{fig9}
  \vspace{-12pt}
\end{figure}

\begin{figure}
  \centering
  \includegraphics[width=0.38\textwidth]{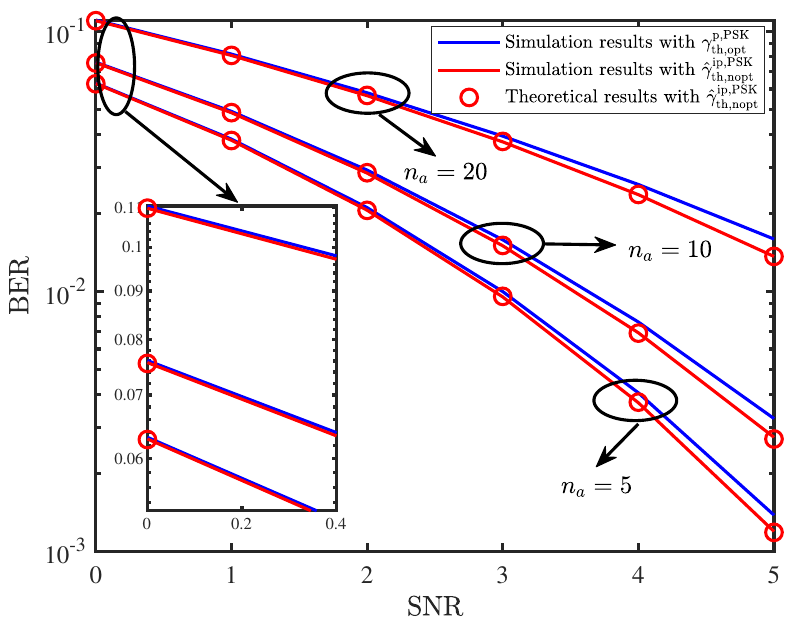}\\
  \centering \caption{\textcolor{black}{BER comparison between ED with the symbol detection threshold $\gamma _{{\rm{th,opt}}}^{{\rm{p,PSK}}}$ in \eqref{25} and that with the symbol detection threshold $\hat \gamma _{{\rm{th,nopt}}}^{{\rm{ip,PSK}}}$ in \ref{28} when PSK signal is used.}}\label{fig10}
  \vspace{-15pt}
\end{figure}

Figs. \ref{fig9} - \ref{fig10} show performances of ED for PSK signal under RTSE. We set $N=100$ for all cases. Again, RTSE degrades the detection performance of ED. Also, our derived near-optimal symbol detection threshold improves BER performance of ED. As $n_a$ increases, effectiveness of $\hat \gamma _{{\rm{th,nopt}}}^{{\rm{ip,PSK}}}$ \ref{28} in reducing BER becomes more pronounced. These observations agree with those for the complex Gaussian signal.

\color{black}

\section{Conclusions}
In this paper, we proposed a new AmBC symbol detection model considering the RTSE between BT and BR. Taking ED as an example, we derived an exact BER expression and an approximate yet concise BER expression, based on which the serious performance degradation caused by RTSE was also highlighted. To improve the detection performance of ED under RTSE, we derived a closed-form expression of the near-optimal symbol detection threshold that requires prior knowledge. We also proposed a novel method to estimate the near-optimal symbol detection threshold by exploiting the attributes of the BR's received samples. Finally, simulation results were provided to verify the theoretical results. In the future, we will study the symbol detection for a AmBC system with multiple BT in the presence of RTSE.

\section*{Appendix A}
\begin{figure*}[t]
\normalsize
\begin{align}\label{A1}\notag
\textcolor{black}{\Gamma _{k|i,j}^{{\rm{ip}}} = N \times \underbrace {\frac{1}{N}\left( {\sum\limits_{n = (k - 1)N + 1}^{(k - 1)N + {n_a}} {{{\left| {y_{k - 1}^{{\rm{ip}}}\left( n \right)|B\left( {k - 1} \right) = i} \right|}^2}}  + \!\!\!\!\!\!\sum\limits_{n = (k - 1)N + {n_a} + 1}^{kN} {{{\left| {y_k^{{\rm{ip}}}\left( n \right)|B\left( k \right) = j} \right|}^2}} } \right)}_{{\rm{average\;power}}},}\tag*{\textcolor{black}{(A.1)}}
\end{align}
\vspace{-15pt}
\hrulefill
\end{figure*}

\renewcommand{\theequation}{A.\arabic{equation}}
Without loss of generality, we take ${\Delta n < 0}$ and ${\sigma _0^2 \le \sigma _1^2}$ as an example to explain why \eqref{8} is valid under RTSE.

Since we assume OOK modulation and equiprobable symbols of the BT, when ${B\left( k \right) = 0}$, ${B\left( {k - 1} \right)}$ can be `0' or `1' with the same probability , i.e., $\Pr \left( {B\left( {k - 1} \right) \!= \!0} \right) \!=\! \Pr \left( {B\left( {k - 1} \right) \!= \!1} \right)\! =\! \frac{1}{2}$. Similarly, when ${B\left( k \right) = 1}$, we also have $\Pr \left( {B\left( {k \!-\! 1} \right) \!= \!0} \right) \!= \!\Pr \left( {B\left( {k - 1} \right) = 1} \right) = \frac{1}{2}$. Using the above results and \ref{A1}, as shown at the top of next page, the average power of the samples, under a sufficiently large $N$, can be written as
\setcounter{equation}{1}
\vspace{-5pt}
\begin{align}\label{A2}
\left\{ {\begin{array}{*{20}{l}}
{\sigma _0^2,}&{{\rm{if}}\;B\left( {k - 1} \right) = 0,B\left( k \right) = 0}\\
{\frac{{{n_a}\sigma _0^2 + \left( {N - {n_a}} \right)\sigma _1^2}}{N},}&{{\rm{if}}\;B\left( {k - 1} \right) = 0,B\left( k \right) = 1}\\
{\frac{{{n_a}\sigma _1^2 + \left( {N - {n_a}} \right)\sigma _0^2}}{N},}&{{\rm{if}}\;B\left( {k - 1} \right) = 1,B\left( k \right) = 0}\\
{\sigma _1^2,}&{{\rm{if}}\;B\left( {k - 1} \right) = 1,B\left( k \right) = 1}
\end{array}} \right..
\end{align}
\vspace{-5pt}

Then, the expectation of the total energy of the $N$ consecutive samples is given as
\vspace{-5pt}
\begin{align}\label{A3}
\left\{ {\begin{array}{*{20}{l}}
{{\mu _{0,0}} \!=\! N\sigma _0^2,}&{{\rm{if}}\;B\left( {k \!-\! 1} \right) \!=\! 0,B\left( k \right) \!=\! 0}\\
{{\mu _{0,1}} \!=\! {n_a}\sigma _0^2 \!+\! \left( {N - {n_a}} \right)\sigma _1^2,}&{{\rm{if}}\;B\left( {k \!-\! 1} \right) \!=\! 0,B\left( k \right) \!=\! 1}\\
{{\mu _{1,0}} \!=\! {n_a}\sigma _1^2 \!+\! \left( {N - {n_a}} \right)\sigma _0^2,}&{{\rm{if}}\;B\left( {k \!-\! 1} \right) \!=\! 1,B\left( k \right) \!=\! 0}\\
{{\mu _{1,1}} \!=\! N\sigma _1^2,}&{{\rm{if}}\;B\left( {k \!-\! 1} \right) \!=\! 1,B\left( k \right) \!=\! 1}
\end{array}} \right..
\end{align}
\vspace{-5pt}

Since this work focuses on the RTSE caused by the imperfect TS, it is reasonable to assume ${n_a} \ll N$ and $\frac{{{n_a}}}{N} < 50\% $, which is verified by the recent experimental findings \cite{10167801} and discussed in \emph{Remark 2}. Therefore, we obtain ${\mu _{0,0}} < {\mu _{1,0}} < {\mu _{0,1}} < {\mu _{1,1}}$ that can be used to distinguish $B(k)$ via the following decision criterion,
\vspace{-5pt}
\begin{align}\label{A4}
\left\{ {\begin{array}{*{20}{c}}
{\hat B\left( k \right) = 0,{\rm{if}}\;\Gamma _k^{{\rm{ip}}} < {n_a}\sigma _0^2 + \left( {N - {n_a}} \right)\sigma _1^2}\\
{\hat B\left( k \right) = 1,{\rm{if}}\;\Gamma _k^{{\rm{ip}}} \ge {n_a}\sigma _0^2 + \left( {N - {n_a}} \right)\sigma _1^2}
\end{array}} \right..
\end{align}
\vspace{-5pt}

However, \eqref{A4} holds for a sufficiently large $N$. For a finite value of $N$ in practical AmBC, $\Gamma _k^{{\rm{ip}}}$ is not a constant; rather, it is presented in the form of a probability density distribution. Thus, the decision criterion should be modified as
\vspace{-5pt}
\begin{align}\label{A5}
\left\{ {\begin{array}{*{20}{c}}
{\hat B\left( k \right) = 0,{\rm{if}}\;\Gamma _k^{{\rm{ip}}} < \gamma _{{\rm{th}}}^{{\rm{ip}}}}\\
{\hat B\left( k \right) = 1,{\rm{if}}\;\Gamma _k^{{\rm{ip}}} \ge \gamma _{{\rm{th}}}^{{\rm{ip}}}}
\end{array}} \right..
\end{align}
\vspace{-5pt}

Similarly, we can obtain the decision criteria under the other three cases, i.e., $\left\{ {\Delta n < 0,\sigma _0^2 > \sigma _1^2} \right\}$, $\left\{ {\Delta n > 0,\sigma _0^2 \le \sigma _1^2} \right\}$, and $\left\{ {\Delta n > 0,\sigma _0^2 > \sigma _1^2} \right\}$, respectively, as follows,
\vspace{-5pt}
\begin{align}\label{A6}
\left\{ {\begin{array}{*{20}{c}}
{\hat B\left( k \right) = 0,{\rm{if}}\;\Gamma _k^{{\rm{ip}}} \ge \gamma _{{\rm{th}}}^{{\rm{ip}}}}\\
{\hat B\left( k \right) = 1,{\rm{if}}\;\Gamma _k^{{\rm{ip}}} < \gamma _{{\rm{th}}}^{{\rm{ip}}}}
\end{array}} \right.,
\end{align}
\vspace{-20pt}

\begin{align}\label{A7}
\left\{ {\begin{array}{*{20}{c}}
{\hat B\left( k \right) = 0,{\rm{if}}\;\Gamma _k^{{\rm{ip}}} < \gamma _{{\rm{th}}}^{{\rm{ip}}}}\\
{\hat B\left( k \right) = 1,{\rm{if}}\;\Gamma _k^{{\rm{ip}}} \ge \gamma _{{\rm{th}}}^{{\rm{ip}}}}
\end{array}} \right.,
\end{align}
\vspace{-20pt}

\begin{align}\label{A8}
\left\{ {\begin{array}{*{20}{c}}
{\hat B\left( k \right) = 0,{\rm{if}}\;\Gamma _k^{{\rm{ip}}} \ge \gamma _{{\rm{th}}}^{{\rm{ip}}}}\\
{\hat B\left( k \right) = 1,{\rm{if}}\;\Gamma _k^{{\rm{ip}}} < \gamma _{{\rm{th}}}^{{\rm{ip}}}}
\end{array}} \right..
\end{align}
\vspace{-5pt}

Using \eqref{A5}, \eqref{A6}, \eqref{A7}, and \eqref{A8}, we obtain the decision criterion under RTSE as \eqref{8} , which is the same as that under perfect TS.

\begin{figure*}[t]
\normalsize
\begin{align}\label{B1}\notag
\Gamma _{k|i,j}^{{\rm{ip}}} &= \sum\limits_{n = (k - 1)N + 1}^{(k - 1)N + {n_a}} {{{\left| {y_{k - 1}^{{\rm{ip}}}\left( n \right)|B\left( {k - 1} \right) = i} \right|}^2}}  + \sum\limits_{n = (k - 1)N + {n_a} + 1}^{kN} {{{\left| {y_k^{{\rm{ip}}}\left( n \right)|B\left( k \right) = j} \right|}^2}} \\ \notag
 &= \left\{ {\begin{array}{*{20}{l}}
{\sum\limits_{n = (k - 1)N + 1}^{kN} {{{\left| {hs(n) + \omega (n)} \right|}^2},} }&{{\rm{if}}\;B\left( {k - 1} \right) = 0,B\left( k \right) = 0}\\
{\sum\limits_{n = (k - 1)N + 1}^{(k - 1)N + {n_a}} {{{\left| {\mu s(n) + \omega (n)} \right|}^2} + \sum\limits_{n = (k - 1)N + {n_a} + 1}^{kN} {{{\left| {hs(n) + \omega (n)} \right|}^2}} } ,}&{{\rm{if}}\;B\left( {k - 1} \right) = 1,B\left( k \right) = 0}\\
{\sum\limits_{n = (k - 1)N + 1}^{kN} {{{\left| {\mu s(n) + \omega (n)} \right|}^2},} }&{{\rm{if}}\;B\left( {k - 1} \right) = 1,B\left( k \right) = 1}\\
{\sum\limits_{n = (k - 1)N + 1}^{(k - 1)N + {n_a}} {{{\left| {hs(n) + \omega (n)} \right|}^2} + \sum\limits_{n = (k - 1)N + {n_a} + 1}^{kN} {{{\left| {\mu s(n) + \omega (n)} \right|}^2}} } ,}&{{\rm{if}}\;B\left( {k - 1} \right) = 0,B\left( k \right) = 1}
\end{array}} \right..\tag{B.1}
\end{align}
\hrulefill
\end{figure*}

\begin{figure*}[t]
\normalsize
\vspace{-15pt}
\begin{align}\label{B8}\notag
{F_{\Gamma _{k|0,1}^{{\rm{ip}}}}}\left( z \right) &= \iint_{y_1>0,y_2>0}f(y_1,y_2)dy_1dy_2 = \int\limits_0^z {d{y_1}} \int\limits_0^{z - {y_1}} {\frac{{{\alpha ^{{n_a}}}{\beta ^{N - {n_a}}}}}{{{2^N}\Gamma \left( {{n_a}} \right)\Gamma \left( {N - {n_a}} \right)}}{e^{ - \frac{{\alpha {y_1} + \beta {y_2}}}{2}}}y_1^{{n_a} - 1}y_2^{N - {n_a} - 1}d{y_2}} \\ \notag
&  = \frac{{{\alpha ^{{n_a}}}\left[ {\Gamma \left( {N - {n_a},0} \right){{\left( {\frac{2}{\alpha }} \right)}^{{n_a}}}\left( {\Gamma \left( {{n_a},0} \right) - \Gamma \left( {{n_a},\frac{{\alpha z}}{2}} \right)} \right) - \int\limits_0^z {{e^{ - \frac{{\alpha {y_1}}}{2}}}y_1^{{n_a} - 1}\Gamma \left( {N - {n_a},\frac{{\beta \left( {z - {y_1}} \right)}}{2}} \right)d{y_1}} } \right]}}{{{2^{{n_a}}}\Gamma \left( {{n_a}} \right)\Gamma \left( {N - {n_a}} \right)}}\\ \notag
&\mathop  = \limits^{(a)} 1 - \frac{{\Gamma \left( {{n_a},\frac{{\alpha z}}{2}} \right)}}{{\Gamma \left( {{n_a}} \right)}} - \frac{{{\alpha ^{{n_a}}}\left( {N - {n_a} - 1} \right)!{e^{ - \frac{{\beta z}}{2}}}}}{{{2^{{n_a}}}\Gamma \left( {{n_a}} \right)\Gamma \left( {N - {n_a}} \right)}}\sum\limits_{m = 0}^{N - {n_a} - 1} {\frac{{{\beta ^m}}}{{{2^m}m!}}\sum\limits_{i = 0}^m {{{\left( { - 1} \right)}^i}C_m^i} {z^{m - i}}\int\limits_0^z {{e^{\frac{{\left( {\beta  - \alpha } \right){y_1}}}{2}}}y_1^{i + {n_a} - 1}d{y_1}} }  \\ \notag
&\mathop  = \limits^{(b)}  = 1 - \frac{{\Gamma \left( {{n_a},\frac{{\alpha z}}{2}} \right)}}{{\Gamma \left( {{n_a}} \right)}} - \frac{{{{\left( {\frac{\alpha }{{\rm{2}}}} \right)}^{{n_a}}}{e^{ - \frac{{\alpha z}}{2}}}}}{{\Gamma \left( {{n_a}} \right)}}\sum\limits_{m = 0}^{N - {n_a} - 1} {\frac{{{\beta ^m}}}{{{2^m}m!}}\sum\limits_{i = 0}^m {{{\left( { - 1} \right)}^i}C_m^i} {z^{m - i}}}  \times \\ \notag
&\left( {\sum\limits_{k = 0}^{i + {n_a} - 1} {\frac{{{{\left( { - 1} \right)}^{^k}}{2^{k + 1}}k!C_{i + {n_a} - 1}^k}}{{{{\left( {\beta  - \alpha } \right)}^{k + 1}}}}{z^{i + {n_a} - 1 - k}}}  - \frac{{{e^{ - \frac{{\left( {\beta  - \alpha } \right)z}}{2}}}{{\left( { - 1} \right)}^{^{i + {n_a} - 1}}}{2^{i + {n_a}}}\left( {i + {n_a} - 1} \right)!}}{{{{\left( {\beta  - \alpha } \right)}^{i + {n_a}}}}}} \right), \tag{B.8}
\end{align}
\vspace{-10pt}
\hrulefill
\end{figure*}

\begin{figure*}[t]
\normalsize
\begin{align}\label{B9}
{{f_{\Gamma _{k|0,1}^{{\rm{ip}}}}}\left( z \right) = \frac{{{\alpha ^{{n_a}}}{\beta ^{N - {n_a}}}{e^{ - \frac{{\alpha z}}{2}}}\sum\limits_{i = 0}^{{n_a} - 1} {C_{{n_a} - 1}^i} {{\left( { - 1} \right)}^i}{z^{{n_a} - 1 - i}}{{\left( {\frac{2}{{\beta  - \alpha }}} \right)}^{N - {n_a} + i}}\left( {\Gamma \left( {N - {n_a} + i,0} \right) - \Gamma \left( {N - {n_a} + i,\frac{{\left( {\beta  - \alpha } \right)z}}{2}} \right)} \right)}}{{{2^N}\Gamma \left( {{n_a}} \right)\Gamma \left( {N - {n_a}} \right)}}}.\tag{B.9}
\end{align}
\vspace{-15pt}
\hrulefill
\end{figure*}

\color{black}
\section*{Appendix B}
\renewcommand{\theequation}{B.\arabic{equation}}

Taking $\Delta n < 0$ as an example, the ED test statistics under RTSE is calculated as \eqref{B1}, as shown at the top of this page.

For these cases with $\left\{ {B\left( {k - 1} \right) = 0,B\left( k \right) = 0} \right\}$ and $\left\{ {B\left( {k - 1} \right) = 1,B\left( k \right) = 1} \right\}$, we can obtain $\frac{{\Gamma _{k|0,0}^{{\rm{ip}}}}}{{\sigma _0^2/2}} \sim {\chi ^2}\left( {2N} \right)$ and $\frac{{\Gamma _{k|1,1}^{{\rm{ip}}}}}{{\sigma _1^2/2}} \sim {\chi ^2}\left( {2N} \right)$, respectively. Then the PDFs of $\Gamma _{k|0,0}^{{\rm{ip}}}$ and $\Gamma _{k|1,1}^{{\rm{ip}}}$ can be derived as, respectively,
\setcounter{equation}{1}
\begin{align} \label{B2}
{f_{\Gamma _{k|0,0}^{{\rm{ip}}}}}\left( z \right) = \frac{\alpha }{{{2^N}\Gamma \left( N \right)}}{e^{ - \frac{{\alpha z}}{2}}}{\left( {\alpha z} \right)^{N - 1}},
\end{align}
\begin{align} \label{B3}
{f_{\Gamma _{k|1,1}^{{\rm{ip}}}}}\left( z \right) = \frac{\beta }{{{2^N}\Gamma \left( N \right)}}{e^{ - \frac{{\beta z}}{2}}}{\left( {\beta z} \right)^{N - 1}},
\end{align}
where $\alpha  = \frac{2}{{\sigma _0^2}}$, $\beta  = \frac{2}{{\sigma _1 ^2}}$, and $\Gamma \left( s \right) = \int\limits_0^\infty  {{t^{s - 1}}} {e^{ - t}}dt$ is the gamma function.

For the case with $\left\{ {B\left( {k - 1} \right) = 0,B\left( k \right) = 1} \right\}$, we decompose the test statistic $\Gamma _{k|0,1}^{{\rm{ip}}}$ into
\begin{align} \label{B4}\notag
\Gamma _{k|0,1}^{{\rm{ip}}} &= \underbrace {\sum\limits_{n = (k - 1)N + 1}^{(k - 1)N + {n_a}} {{{\left| {hs(n) + \omega (n)} \right|}^2}} }_{{Y_1}}\\ \notag
 &+ \underbrace {\sum\limits_{n = (k - 1)N + {n_a} + 1}^{kN} {{{\left| {\mu s(n) + \omega (n)} \right|}^2}} }_{{Y_2}}. \tag{B.4}
\end{align}

\begin{figure*}[t]
\normalsize
\begin{align}\label{B10}
{{f_{\Gamma _{k|1,0}^{{\rm{ip}}}}}\left( z \right) = \frac{{{\alpha ^{N - {n_a}}}{\beta ^{{n_a}}}{e^{ - \frac{{\beta z}}{2}}}\sum\limits_{i = 0}^{{n_a} - 1} {C_{{n_a} - 1}^i} {{\left( { - 1} \right)}^i}{z^{{n_a} - 1 - i}}{{\left( {\frac{{ - 2}}{{\beta  - \alpha }}} \right)}^{N - {n_a} + i}}\left( {\Gamma \left( {N - {n_a} + i,0} \right) - \Gamma \left( {N - {n_a} + i,\frac{{ - \left( {\beta  - \alpha } \right)z}}{2}} \right)} \right)}}{{{2^N}\Gamma \left( {{n_a}} \right)\Gamma \left( {N - {n_a}} \right)}}}.\tag{B.10}
\end{align}
\vspace{-15pt}
\hrulefill
\end{figure*}

\begin{figure*}[t]
\normalsize
\begin{align}\label{B11} \notag
P_{{\rm{BER|}}\sigma _0^2 > \sigma _1^2}^{{\rm{ip}}} &= \Pr \left( {B\left( {k - 1} \right) = 0,B\left( k \right) = 0} \right)\Pr \left( {\Gamma _{k|0,0}^{{\rm{ip}}} < \gamma _{{\rm{th}}}^{{\rm{ip}}}|B\left( {k - 1} \right) = 0,B\left( k \right) = 0} \right)\\ \notag
 &+ \Pr \left( {B\left( {k - 1} \right) = 1,B\left( k \right) = 0} \right)\Pr \left( {\Gamma _{k|1,0}^{{\rm{ip}}} < \gamma _{{\rm{th}}}^{{\rm{ip}}}|B\left( {k - 1} \right) = 1,B\left( k \right) = 0} \right)\\ \notag
 &+ \Pr \left( {B\left( {k - 1} \right) = 1,B\left( k \right) = 1} \right)\Pr \left( {\Gamma _{k|1,1}^{{\rm{ip}}} \ge \gamma _{{\rm{th}}}^{{\rm{ip}}}|B\left( {k - 1} \right) = 1,B\left( k \right) = 1} \right)\\ \notag
 &+ \Pr \left( {B\left( {k - 1} \right) = 0,B\left( k \right) = 1} \right)\Pr \left( {\Gamma _{k|0,1}^{{\rm{ip}}} \ge \gamma _{{\rm{th}}}^{{\rm{ip}}}|B\left( {k - 1} \right) = 0,B\left( k \right) = 1} \right)\\ \notag
 &= \frac{1}{4}\int\limits_{ - \infty }^{\gamma _{{\rm{th}}}^{{\rm{ip}}}} {{f_{\Gamma _{k|0,0}^{{\rm{ip}}}}}\left( z \right)dz}  + \frac{1}{4}\int\limits_{ - \infty }^{\gamma _{{\rm{th}}}^{{\rm{ip}}}} {{f_{\Gamma _{k|1,0}^{{\rm{ip}}}}}\left( z \right)dz}  + \frac{1}{4}\int\limits_{\gamma _{{\rm{th}}}^{{\rm{ip}}}}^\infty  {{f_{\Gamma _{k|1,1}^{{\rm{ip}}}}}\left( z \right)dz}  + \frac{1}{4}\int\limits_{\gamma _{{\rm{th}}}^{{\rm{ip}}}}^\infty  {{f_{\Gamma _{k|0,1}^{{\rm{ip}}}}}\left( z \right)dz}.\tag{B.11}
\end{align}
\hrulefill
\end{figure*}

\setcounter{equation}{11}
\begin{figure*}[t]
\normalsize
\vspace{-15pt}
\begin{align}\label{B12} \notag
P_{{\rm{BER|}}\sigma _0^2 \le \sigma _1^2}^{{\rm{ip}}} &= \Pr \left( {B\left( {k - 1} \right) = 0,B\left( k \right) = 0} \right)\Pr \left( {\Gamma _{k|0,0}^{{\rm{ip}}} \ge \gamma _{{\rm{th}}}^{{\rm{ip}}}|B\left( {k - 1} \right) = 0,B\left( k \right) = 0} \right)\\ \notag
 &+ \Pr \left( {B\left( {k - 1} \right) = 1,B\left( k \right) = 0} \right)\Pr \left( {\Gamma _{k|1,0}^{{\rm{ip}}} \ge \gamma _{{\rm{th}}}^{{\rm{ip}}}|B\left( {k - 1} \right) = 1,B\left( k \right) = 0} \right)\\ \notag
 &+ \Pr \left( {B\left( {k - 1} \right) = 1,B\left( k \right) = 1} \right)\Pr \left( {\Gamma _{k|1,1}^{{\rm{ip}}} < \gamma _{{\rm{th}}}^{{\rm{ip}}}|B\left( {k - 1} \right) = 1,B\left( k \right) = 1} \right)\\ \notag
 &+ \Pr \left( {B\left( {k - 1} \right) = 0,B\left( k \right) = 1} \right)\Pr \left( {\Gamma _{k|0,1}^{{\rm{ip}}} < \gamma _{{\rm{th}}}^{{\rm{ip}}}|B\left( {k - 1} \right) = 0,B\left( k \right) = 1} \right)\\ \notag
 &= \frac{1}{4}\int\limits_{\gamma _{{\rm{th}}}^{{\rm{ip}}}}^\infty  {{f_{\Gamma _{k|0,0}^{{\rm{ip}}}}}\left( z \right)dz}  + \frac{1}{4}\int\limits_{\gamma _{{\rm{th}}}^{{\rm{ip}}}}^\infty  {{f_{\Gamma _{k|1,0}^{{\rm{ip}}}}}\left( z \right)dz}  + \frac{1}{4}\int\limits_{ - \infty }^{\gamma _{{\rm{th}}}^{{\rm{ip}}}} {{f_{\Gamma _{k|1,1}^{{\rm{ip}}}}}\left( z \right)dz}  + \frac{1}{4}\int\limits_{ - \infty }^{\gamma _{{\rm{th}}}^{{\rm{ip}}}} {{f_{\Gamma _{k|0,1}^{{\rm{ip}}}}}\left( z \right)dz}. \tag{B.12}
\end{align}
\vspace{-15pt}
\hrulefill
\end{figure*}

Then, the PDFs of $Y_1$ and $Y_2$ can be derived as, respectively,
\setcounter{equation}{4}
\begin{align} \label{B5}
{f_{{Y_1}}}\left( {{y_1}} \right) = \frac{\alpha }{{{2^{{n_a}}}\Gamma \left( {{n_a}} \right)}}{e^{ - \frac{{\alpha {y_1}}}{2}}}{\left( {\alpha {y_1}} \right)^{{n_a} - 1}},
\end{align}

\begin{align} \label{B6}
{f_{{Y_2}}}\left( {{y_2}} \right) = \frac{\beta }{{{2^{N - {n_a}}}\Gamma \left( {N - {n_a}} \right)}}{e^{ - \frac{{\beta {y_2}}}{2}}}{\left( {\beta {y_2}} \right)^{N - {n_a} - 1}}.
\end{align}

Then the joint PDF of variables $Y_1$ and $Y_2$ is given by
\begin{align} \label{B7}
f\left( {{y_1},{y_2}} \right) = \frac{{{\alpha ^{{n_a}}}{\beta ^{N \!\!-\!\! {n_a}}}}}{{{2^N}\Gamma \left( {{n_a}} \right)\Gamma \left( {N - {n_a}} \right)}}{e^{ - \frac{{\alpha {y_1} + \beta {y_2}}}{2}}}y_1^{{n_a} - 1}y_2^{N - {n_a} - 1}.
\end{align}

The distribution function of $\Gamma _{k|0,1}^{{\rm{ip}}}$ can be expressed as \eqref{B8}, as shown at the top of the previous page, where steps (a) and (b) are derived from eqs. (8.352.4) and (2.321.2) in \cite{zwillinger2007table}, respectively, and $\Gamma \left( {s,x} \right) = \int\limits_x^\infty  {{t^{s - 1}}} {e^{ - t}}dt$ is the upper incomplete gamma function.

By taking the derivative of \eqref{B8}, we can get the PDF of $\Gamma _{k|0,1}^{{\rm{ip}}}$ as \eqref{B9}, as shown at the top of the previous page.  Similarly, we can get PDF of $\Gamma _{k|1,0}^{{\rm{ip}}}$ as \eqref{B10}, as shown at the top of this page.

Using \eqref{B2}, \eqref{B3}, \eqref{B9}, and \eqref{B10}, for a given symbol detection threshold ${\gamma _{{\rm{th}}}^{{\rm{ip}}}}$ and RTSE $n_a$, the BERs under $\sigma _0^2 > \sigma _1^2$ and $\sigma _0^2 \le \sigma _1^2$ can be derived as \eqref{B11} and \eqref{B12}, respectively, as shown at the top of this page. Thus, when $\Delta n < 0$, we can obtain the BER expression as \eqref{10}. Similar as above, when $\Delta n > 0$, we can derive the BER expression, same as in \eqref{10}.

\color{black}
\section*{Appendix C}
\begin{figure*}[t]
\normalsize
\begin{align} \label{C4} \notag
&\sum\limits_{n = \left( {k - 1} \right)N + 1}^{kN} {\mathbb{E}\left[ {{{\left| {{{\left| {y_k^{{\rm{ip}}}\left( n \right)} \right|}^2}\! - \!\mathbb{E}\left[ {{{\left| {y_k^{{\rm{ip}}}\left( n \right)} \right|}^2}} \right]} \right|}^{2 + \delta }}} \right]}
 \mathop  = \limits^{(c)} \sum\limits_{n = \left( {k - 1} \right)N + 1}^{\left( {k - 1} \right)N + {n_a}} {\mathbb{E}\left[ {{{\left| {{Y_{1,n}} \!-\! \mathbb{E}\left[ {{Y_{1,n}}} \right]} \right|}^3}} \right]}
 \!+\! \sum\limits_{n = \left( {k - 1} \right)N + {n_a} + 1}^{kN} \! {\mathbb{E}\left[ {{{\left| {{Y_{2,n}} \!-\! \mathbb{E}\left[ {{Y_{2,n}}} \right]} \right|}^3}} \right]} \\ \notag
 = &{n_a}\left( {\mathbb{E}\left[ {Y_{1,n}^3} \right] - 3\mathbb{E}\left[ {{Y_{1,n}}} \right]\mathbb{E}\left[ {Y_{_{1,n}}^2} \right] + 2{{\left( {\mathbb{E}\left[ {{Y_{1,n}}} \right]} \right)}^3}} \right)
 + \left( {N - {n_a}} \right)\left( {\mathbb{E}\left[ {Y_{2,n}^3} \right] - 3\mathbb{E}\left[ {{Y_{2,n}}} \right]\mathbb{E}\left[ {Y_{_{2,n}}^2} \right] + 2{{\left( {\mathbb{E}\left[ {{Y_{2,n}}} \right]} \right)}^3}} \right)\\ \notag
 = &\frac{{16{n_a}}}{{{\alpha ^3}}} + \frac{{16\left( {N - {n_a}} \right)}}{{{\beta ^3}}}, \tag{C.4}
\end{align}
\vspace{-15pt}
\hrulefill
\end{figure*}

\begin{figure*}[t]
\normalsize
\begin{align}\label{E6}
\textcolor{black}{
\left\{ {\begin{array}{*{20}{l}}
{\Pr \left( {\hat B\left( k \right) = 1|B\left( k \right) = 0} \right) = Q\left( {\frac{{\gamma _{{\rm{th,ML}}}^{{\rm{ip0}}} - N\sigma _0^2}}{{\sqrt {N\sigma _0^4} }}{\mathop{\rm sgn}} \left( {\sigma _1^2 - \sigma _0^2} \right)} \right)}\\
{\Pr \left( {\hat B\left( k \right) = 0|B\left( k \right) = 1} \right) = Q\left( {\frac{{\left( {{n_a}\sigma _0^2 + \left( {N - {n_a}} \right)\sigma _1^2} \right) - \gamma _{{\rm{th,ML}}}^{{\rm{ip0}}}}}{{\sqrt {{n_a}\sigma _0^4 + \left( {N - {n_a}} \right)\sigma _1^4} }}{\mathop{\rm sgn}} \left( {\sigma _1^2 - \sigma _0^2} \right)} \right)}
\end{array}} \right.,}\tag*{\textcolor{black}{(E.6)}}
\end{align}
\hrulefill
\end{figure*}

\begin{figure*}[t]
\normalsize
\vspace{-15pt}
\begin{align}\label{E7}
\textcolor{black}{
\left\{ {\begin{array}{*{20}{l}}
{\Pr \left( {\hat B\left( k \right) = 1|B\left( k \right) = 0} \right) = Q\left( {\frac{{\gamma _{{\rm{th,ML}}}^{{\rm{ip1}}} - \left( {{n_a}\sigma _1^2 + \left( {N - {n_a}} \right)\sigma _0^2} \right)}}{{\sqrt {{n_a}\sigma _1^4 + \left( {N - {n_a}} \right)\sigma _0^4} }}{\mathop{\rm sgn}} \left( {\sigma _1^2 - \sigma _0^2} \right)} \right)}\\
{\Pr \left( {\hat B\left( k \right) = 0|B\left( k \right) = 1} \right) = Q\left( {\frac{{N\sigma _1^2 - \gamma _{{\rm{th,ML}}}^{{\rm{ip1}}}}}{{\sqrt {N\sigma _1^4} }}{\mathop{\rm sgn}} \left( {\sigma _1^2 - \sigma _0^2} \right)} \right)}
\end{array}} \right.,}\tag*{\textcolor{black}{(E.7)}}
\end{align}
\vspace{-10pt}
\hrulefill
\end{figure*}

\begin{figure*}[t]
\normalsize
\begin{align}\label{E8}
\textcolor{black}{
Q\left( {\frac{{\gamma _{{\rm{th,ML}}}^{{\rm{ip0}}} - N\sigma _0^2}}{{\sqrt {N\sigma _0^4} }}{\mathop{\rm sgn}} \left( {\sigma _1^2 - \sigma _0^2} \right)} \right) \ne Q\left( {\frac{{\left( {{n_a}\sigma _0^2 + \left( {N - {n_a}} \right)\sigma _1^2} \right) - \gamma _{{\rm{th,ML}}}^{{\rm{ip0}}}}}{{\sqrt {{n_a}\sigma _0^4 + \left( {N - {n_a}} \right)\sigma _1^4} }}{\mathop{\rm sgn}} \left( {\sigma _1^2 - \sigma _0^2} \right)} \right),}\tag*{\textcolor{black}{(E.8)}}
\end{align}
\vspace{-15pt}
\hrulefill
\end{figure*}

\renewcommand{\theequation}{C.\arabic{equation}}
When $\Delta n <0 $, taking the case with $\left\{ {B\left( {k - 1} \right) = 0,B\left( k \right) = 1} \right\}$ as an example, we decompose the test statistic $\Gamma _{k|0,1}^{{\rm{ip}}}$ into
\begin{align} \label{C1}\notag
\Gamma _{k|0,1}^{{\rm{ip}}} &= \sum\limits_{n = \left( {k - 1} \right)N + 1}^{\left( {k - 1} \right)N + {n_a}} {\underbrace {{{\left| {hs(n) + \omega (n)} \right|}^2}}_{{Y_{1,n}}}} \\ \notag
 &+ \sum\limits_{n = \left( {k - 1} \right)N + {n_a} + 1}^{kN} {\underbrace {{{\left| {\mu s(n) + \omega (n)} \right|}^2}}_{{Y_{2,n}}}}.\tag{C.1}
\end{align}

It is easy to verify $\frac{{{Y_{1,n}}}}{{\sigma _0^2/2}} \sim {\chi ^2}\left( {2} \right)$ and $\frac{{{Y_{2,n}}}}{{\sigma _1^2/2}} \sim {\chi ^2}\left( {2} \right)$, where ${\chi ^2}\left( m \right)$ represents a chi-square distribution with $m$ degrees of freedom. Next, we can easily know
\setcounter{equation}{1}
\begin{align} \label{C2}
\left\{ {\begin{array}{*{20}{c}}
{\mathbb{E} \left[ {{Y_{1,n}}} \right] = \frac{2}{\alpha },\mathbb{D} \left[ {{Y_{1,n}}} \right] = \frac{4}{{{\alpha ^2}}}}\\
{\mathbb{E} \left[ {{Y_{2,n}}} \right] = \frac{2}{\beta },\mathbb{D} \left[ {{Y_{2,n}}} \right] = \frac{4}{{{\beta ^2}}}}
\end{array}} \right..
\end{align}

Therefore, we can get
\begin{align} \label{C3} \notag
{s_{N,k}} &= \sqrt {\sum\limits_{n = (k - 1)N + 1}^{kN} {\mathbb{D}\left[ {{{\left| {y_k^{{\rm{ip}}}\left( n \right)} \right|}^2}} \right]} } \\ \notag
 &= \sqrt {\sum\limits_{n = \left( {k - 1} \right)N + 1}^{\left( {k - 1} \right)N + {n_a}} {\mathbb{D}\left[ {{Y_{1,n}}} \right]}  + \sum\limits_{n = \left( {k - 1} \right)N + {n_a} + 1}^{kN} {\mathbb{D}\left[ {{Y_{2,n}}} \right]} } \\ \notag
 &= \sqrt {\frac{{4{n_a}}}{{{\alpha ^2}}} + \frac{{4\left( {N - {n_a}} \right)}}{{{\beta ^2}}}}.\tag{C.3}
\end{align}

The characteristic function of ${{Y_{1,n}}}$ and ${{Y_{2,n}}}$ are ${\varphi _{{Y_{1,n}}}}\left( t \right) = \frac{\alpha }{{\alpha  - 2it}}$ and ${\varphi _{{Y_{2,n}}}}\left( t \right) = \frac{\beta }{{\beta  - 2it}}$, respectively, where $i$ denotes the imaginary number. According to the property of the characteristic function, we can obtain $\mathbb{E}\left[ {Y_{_{1,n}}^2} \right] = \frac{8}{{{\alpha ^2}}}$, $\mathbb{E}\left[ {Y_{_{1,n}}^3} \right] = \frac{{48}}{{{\alpha ^3}}}$, $\mathbb{E}\left[ {Y_{_{2,n}}^2} \right] = \frac{8}{{{\beta ^2}}}$ and $\mathbb{E}\left[ {Y_{_{2,n}}^3} \right] = \frac{{48}}{{{\beta ^3}}}$. Using these results, we can get \eqref{C4}, as shown at the top of this page, where $\delta$ is a positive constant chosen at random, and $\delta \!\!=\!\! 1$ in step (c).

Therefore, the existence of a positive constant $\delta$ ensures
\setcounter{equation}{4}
\begin{align} \label{C5}
\mathop {\lim }\limits_{N \to \infty } \!\!\!\frac{{\sum\limits_{n = \left( {k \!-\! 1} \right)N \!+\! 1}^{kN} \!\!\!\!{\mathbb{E}\left[ {{{\left| {{{\left| {y_k^{{\rm{ip}}}\left( n \right)} \right|}^2}\!\!\! - \!\! \mathbb{E}\left[ {{{\left| {y_k^{{\rm{ip}}}\left( n \right)} \right|}^2}} \right]} \right|}^{2 + \delta }}} \right]} }}{{{{\left( {{s_{N,k}}} \right)}^{2 + \delta }}}} \!=\! 0.
\end{align}

Similarly, it can be proved in other different cases, and the proof of Lemma 1 is complete.

\section*{Appendix D}
\renewcommand{\theequation}{D.\arabic{equation}}
For a given symbol detection threshold ${\gamma _{{\rm{th}}}^{{\rm{ip}}}}$ and the RTSE $n_a$, according to the approximate PDF \eqref{12} of $\Gamma _{k|i,j}^{{\rm{ip}}}$, the BER achieved by the ED can be derived as follows.

1) If $\sigma _0^2 > \sigma _1^2$, the BER is given as
\begin{align}\label{D1}\notag
P_{{\rm{BER|}}\sigma _0^2 > \sigma _1^2}^{{\rm{ip}}} &= \frac{1}{4}\int\limits_{ - \infty }^{\gamma _{{\rm{th}}}^{{\rm{ip}}}} {{f_{\Gamma _{k|0,0}^{{\rm{ip}}}}}\left( z \right)dz}  + \frac{1}{4}\int\limits_{ - \infty }^{\gamma _{{\rm{th}}}^{{\rm{ip}}}} {{f_{\Gamma _{k|1,0}^{{\rm{ip}}}}}\left( z \right)dz} \\ \notag
 &+ \frac{1}{4}\int\limits_{\gamma _{{\rm{th}}}^{{\rm{ip}}}}^\infty  {{f_{\Gamma _{k|1,1}^{{\rm{ip}}}}}\left( z \right)dz}  + \frac{1}{4}\int\limits_{\gamma _{{\rm{th}}}^{{\rm{ip}}}}^\infty  {{f_{\Gamma _{k|0,1}^{{\rm{ip}}}}}\left( z \right)dz} \\ \notag
 & \simeq  \frac{1}{4}Q\left( {\frac{{N\sigma _0^2 - \gamma _{{\rm{th}}}^{{\rm{ip}}}}}{{\sqrt N \sigma _0^2}}} \right) + \frac{1}{4}Q\left( {\frac{{\gamma _{{\rm{th}}}^{{\rm{ip}}} - N\sigma _1^2}}{{\sqrt N \sigma _1^2}}} \right)\\ \notag
 &+ \frac{1}{4}Q\left( {\frac{{\left( {{n_a}\sigma _1^2 \!+\! \left( {N \!-\! {n_a}} \right)\sigma _0^2} \right) \!-\! \gamma _{{\rm{th}}}^{{\rm{ip}}}}}{{\sqrt {{n_a}\sigma _1^4 + \left( {N - {n_a}} \right)\sigma _0^4} }}} \right)\\ \notag
 &+ \frac{1}{4}Q\left( {\frac{{\gamma _{{\rm{th}}}^{{\rm{ip}}} \!-\! \left( {{n_a}\sigma _0^2 \!+\! \left( {N \!-\! {n_a}} \right)\sigma _1^2} \right)}}{{\sqrt {{n_a}\sigma _0^4 + \left( {N - {n_a}} \right)\sigma _1^4} }}} \right) . \tag{D.1}
\end{align}

2) If $\sigma _0^2  \le \sigma _1^2$, the BER is given as
\begin{align}\label{D2}\notag
P_{{\rm{BER|}}\sigma _0^2 \le \sigma _1^2}^{{\rm{ip}}} &= \frac{1}{4}\int\limits_{\gamma _{{\rm{th}}}^{{\rm{ip}}}}^\infty  {{f_{\Gamma _{k|0,0}^{{\rm{ip}}}}}\left( z \right)dz}  + \frac{1}{4}\int\limits_{\gamma _{{\rm{th}}}^{{\rm{ip}}}}^\infty  {{f_{\Gamma _{k|1,0}^{{\rm{ip}}}}}\left( z \right)dz} \\ \notag
 &+ \frac{1}{4}\int\limits_{ - \infty }^{\gamma _{{\rm{th}}}^{{\rm{ip}}}} {{f_{\Gamma _{k|1,1}^{{\rm{ip}}}}}\left( z \right)dz}  + \frac{1}{4}\int\limits_{ - \infty }^{\gamma _{{\rm{th}}}^{{\rm{ip}}}} {{f_{\Gamma _{k|0,1}^{{\rm{ip}}}}}\left( z \right)dz} \\ \notag
 &\simeq \frac{1}{4}Q\left( {\frac{{\gamma _{{\rm{th}}}^{{\rm{ip}}} - N\sigma _0^2}}{{\sqrt N \sigma _0^2}}} \right) + \frac{1}{4}Q\left( {\frac{{N\sigma _1^2 - \gamma _{{\rm{th}}}^{{\rm{ip}}}}}{{\sqrt N \sigma _1^2}}} \right)\\ \notag
 &+ \frac{1}{4}Q\left( {\frac{{\gamma _{{\rm{th}}}^{{\rm{ip}}} \!-\! \left( {{n_a}\sigma _1^2 \!+\! \left( {N \!-\! {n_a}} \right)\sigma _0^2} \right)}}{{\sqrt {{n_a}\sigma _1^4 + \left( {N - {n_a}} \right)\sigma _0^4} }}} \right)\\ \notag
 &+ \frac{1}{4}Q\left( {\frac{{\left( {{n_a}\sigma _0^2 \!+\! \left( {N \!-\! {n_a}} \right)\sigma _1^2} \right) \!-\! \gamma _{{\rm{th}}}^{{\rm{ip}}}}}{{\sqrt {{n_a}\sigma _0^4 + \left( {N - {n_a}} \right)\sigma _1^4} }}} \right).\tag{D.2}
\end{align}

Therefore, substituting \eqref{D1} and \eqref{D2} into the total probability formula of $P_{{\rm{BER}}}^{{\rm{ip}}}$, \eqref{13} can be obtained and the proof of Theorem 2 is complete.

\color{black}
\section*{Appendix E}

\begin{figure*}[t]
\normalsize
\begin{align}\label{E9}
\textcolor{black}{
Q\left( {\frac{{N\sigma _1^2 - \gamma _{{\rm{th,ML}}}^{{\rm{ip1}}}}}{{\sqrt {N\sigma _1^4} }}{\mathop{\rm sgn}} \left( {\sigma _1^2 - \sigma _0^2} \right)} \right) \ne Q\left( {\frac{{\gamma _{{\rm{th,ML}}}^{{\rm{ip1}}} - \left( {{n_a}\sigma _1^2 + \left( {N - {n_a}} \right)\sigma _0^2} \right)}}{{\sqrt {{n_a}\sigma _1^4 + \left( {N - {n_a}} \right)\sigma _0^4} }}{\mathop{\rm sgn}} \left( {\sigma _1^2 - \sigma _0^2} \right)} \right),}\tag*{\textcolor{black}{(E.9)}}
\end{align}
\vspace{-15pt}
\hrulefill
\end{figure*}

\begin{figure*}[t]
\normalsize
\begin{align}\label{G1}
\frac{{\gamma _{{\rm{th,nopt}}}^{{\rm{ip}}} \!\!-\!\! N\sigma _{\min }^2}}{{\sqrt N \sigma _{\min }^2}} \!\!=\!\! \frac{1}{2}\frac{{\left( {N - {n_a}} \right)\left( {\sigma _{\max }^2 - \sigma _{\min }^2} \right)}}{{\left( {\sqrt N \sigma _{\min }^2 \!\!+\!\! \sqrt {{n_a}\sigma _{\min }^4 \!\!+\!\! \left( {N \!-\! {n_a}} \right)\sigma _{\max }^4} } \right)}}
\!\!+\!\! \frac{1}{2}\frac{{\left( {\sigma _{\max }^2 \!\!-\!\! \sigma _{\min }^2} \right)\left( {\sqrt N \sqrt {{n_a}\sigma _{\max }^4 \!\!+\!\! \left( {N \!-\! {n_a}} \right)\sigma _{\min }^4}  \!\!- \! {n_a}\sigma _{\max }^2} \right)}}{{\sigma _{\min }^2\left( {\sqrt {{n_a}\sigma _{\max }^4 + \left( {N - {n_a}} \right)\sigma _{\min }^4}  + \sqrt N \sigma _{\max }^2} \right)}}.\tag{G.1}
\end{align}
\vspace{-20pt}
\hrulefill
\end{figure*}

\renewcommand{\theequation}{E.\arabic{equation}}

Here we explain why ML criterion \cite{proakis2008digital} leads to an unbalanced BER under RTSE by a mathematical analysis by  taking $\Delta n < 0$ as an example.

For the case with $B\left( {k - 1} \right) = 0$, the optimal symbol detection threshold obtained using ML, denoted by ${\gamma _{{\rm{th,ML}}}^{{\rm{ip0}}}}$, is derived. Similarly, for the case with $B\left( {k - 1} \right) = 1$, the optimal symbol detection threshold obtained using ML, denoted by ${\gamma _{{\rm{th,ML}}}^{{\rm{ip1}}}}$, is derived. Both ${\gamma _{{\rm{th,ML}}}^{{\rm{ip0}}}}$ and ${\gamma _{{\rm{th,ML}}}^{{\rm{ip1}}}}$ can be determined by solving
\setcounter{equation}{0}
\begin{align}\label{E1}
{f_{\Gamma _{k|i,0}^{{\rm{ip}}}}}\left( z \right) = {f_{\Gamma _{k|i,1}^{{\rm{ip}}}}}\left( z \right){|_{z = \gamma _{{\rm{th,ML}}}^{{\rm{ip}}}}},
\end{align}
where $\gamma _{{\rm{th,ML}}}^{{\rm{ip}}} = \left( {1 - i} \right)\gamma _{{\rm{th,ML}}}^{{\rm{ip0}}} + i\gamma _{{\rm{th,ML}}}^{{\rm{ip1}}}$, $B\left( {k - 1} \right) = i$, and $i \in \left\{ {0,1} \right\}$.

Using $\Gamma _{k|i,j}^{{\rm{ip}}} \sim \mathbb{N}\left( {{\mu _{i,j}},{\varsigma _{i,j}}} \right)$, we can obtain the following equation, given by
\begin{align}\label{E2}\notag
\frac{{\exp \left[ { - \frac{{{{\left( {z - {\mu _{i,0}}} \right)}^2}}}{{2{\varsigma _{i,0}}}}} \right]}}{{\sqrt {2\pi {\varsigma _{i,0}}} }} = \frac{{\exp \left[ { - \frac{{{{\left( {z - {\mu _{i,1}}} \right)}^2}}}{{2{\varsigma _{i,1}}}}} \right]}}{{\sqrt {2\pi {\varsigma _{i,1}}} }}{|_{z = \gamma _{{\rm{th,ML}}}^{{\rm{ip}}}}}.\tag{E.2}
\end{align}

After taking the logarithms of both sides of  \eqref{E2}, the resulting expression can be further simplified as
\begin{align}\label{E3}\notag
&\frac{{C_i^ +  - 1}}{2}{\left( {\gamma _{{\rm{th,ML}}}^{{\rm{ip}}}} \right)^2} + \left( {{\mu _{i,1}} - {\mu _{i,0}}C_i^ + } \right)\gamma _{{\rm{th,ML}}}^{{\rm{ip}}} \\ \notag
+& \frac{{\mu _{i,0}^2C_i^ +  - \mu _{i,1}^2 - {\varsigma _{i,1}}\ln C_i^ + }}{2} = 0,\tag{E.3}
\end{align}
where $C_i^ +  = \frac{{{\varsigma _{i,1}}}}{{{\varsigma _{i,0}}}}$.

After mathematical transformations, both ${\gamma _{{\rm{th,ML}}}^{{\rm{ip0}}}}$ and ${\gamma _{{\rm{th,ML}}}^{{\rm{ip1}}}}$ can be written as, respectively,
\setcounter{equation}{3}
\begin{align}\label{E4}
\gamma _{{\rm{th,ML}}}^{{\rm{ip0}}} = \frac{{{{\bar \mu }_0} \!+\! \sqrt {C_0^ + {{\left( {{\mu _{0,1}} \!-\! {\mu _{0,0}}} \right)}^2} + C_0^ + \left( {{\varsigma _{0,1}}\!-\! {\varsigma _{0,0}}} \right)\ln C_0^ + } }}{{C_0^ +  - 1}},
\end{align}

\begin{align}\label{E5}
\gamma _{{\rm{th,ML}}}^{{\rm{ip1}}} = \frac{{{{\bar \mu }_1} \!+\! \sqrt {C_1^ + {{\left( {{\mu _{1,1}} \!-\! {\mu _{1,0}}} \right)}^2} + C_1^ + \left( {{\varsigma _{1,1}} \!-\! {\varsigma _{1,0}}} \right)\ln C_1^ + } }}{{C_1^ +  - 1}},
\end{align}
where ${{\bar \mu }_0} = {\mu _{0,0}}C_0^ +  - {\mu _{0,1}}$ and ${{\bar \mu }_1} = {\mu _{1,0}}C_1^ +  - {\mu _{1,1}}$.

Based on the BER expression, ${\gamma _{{\rm{th,ML}}}^{{\rm{ip0}}}}$ and ${\gamma _{{\rm{th,ML}}}^{{\rm{ip1}}}}$ in  \eqref{E4} and \eqref{E5}, for the case with $B\left( {k - 1} \right) = 0$, the BER is given by \ref{E6}, as shown at the top of the previous page, where ${\mathop{\rm sgn}} \left( {\sigma _1^2 - \sigma _0^2} \right) = \left\{ {\begin{array}{*{20}{c}}
{1,{\rm{if}}\;\sigma _1^2 - \sigma _0^2 > 0}\\
{0,{\rm{if}}\;\sigma _1^2 - \sigma _0^2 = 0}\\
{ - 1,{\rm{if}}\;\sigma _1^2 - \sigma _0^2 < 0}
\end{array}} \right.$.

Similarly, for the case with $B\left( {k - 1} \right) = 1$, the BER is given by \ref{E7}, as shown at the top of the previous page.

After mathematical transformations, we can obtain \ref{E8} and \ref{E9}, as shown at the top of the previous and this page, respectively.

\emph{Remark E.} If $\sigma _0^2 = \sigma _1^2$, a balanced BER is achieved under RTSE. However, there is no need to consider the scenario where $\sigma _0^2 = \sigma _1^2$ \cite{8007328}. Therefore, an unbalanced BER is achieved under RTSE.

Similarly, we can obtain the same conclusion when $\Delta n > 0$.

\section*{Appendix F}
\renewcommand{\theequation}{F.\arabic{equation}}
Taking $\Delta n < 0$ as an example, according to the received signal samples and estimated $\sigma _{\min }^2$ and $\sigma _{\max }^2$, the RTSE $n_a$ can be obtained as follows.

1) If $\sigma _0^2 > \sigma _1^2$, we can obtain
\setcounter{equation}{0}
\begin{align}\label{F1}
{E_{10}} - {E_{01}} = \frac{{\left( {N - 2{n_a}} \right)\left( {\sigma _0^2 - \sigma _1^2} \right)}}{N}.
\end{align}

After mathematical transformations and using $\frac{{{n_a}}}{N} < 50\%$, $n_a$ is given by
\vspace{-5pt}
\begin{align}\label{F2}
{n_a} = \frac{N}{2}\left( {1 - \frac{{{E_{10}} - {E_{01}}}}{{\sigma _0^2 - \sigma _1^2}}} \right).
\end{align}

2) If $\sigma _0^2  \le \sigma _1^2$, we can obtain
\begin{align}\label{F3}
{E_{01}} - {E_{10}} = \frac{{\left( {N - 2{n_a}} \right)\left( {\sigma _1^2 - \sigma _0^2} \right)}}{N}.
\end{align}

After mathematical transformations and using $\frac{{{n_a}}}{N} < 50\%$, $n_a$ is given by
\vspace{-5pt}
\begin{align}\label{F4}
{n_a} = \frac{N}{2}\left( {1 - \frac{{{E_{01}} - {E_{10}}}}{{\sigma _1^2 - \sigma _0^2}}} \right).
\end{align}

Therefore, substituting \eqref{F2} and \eqref{F4} into the total expression of the RTSE $n_a$, we can obtain
\begin{align}\label{F5}
{\hat n_a} = \frac{N}{2}\left( {1 - \frac{{{E_3} - {E_2}}}{{\sigma _{\max }^2 - \sigma _{\min }^2}}} \right).
\end{align}

Similar as above, when $\Delta n > 0$, we can derive an expression of the RTSE $\hat n_a$, same as in \eqref{F5}.

\section*{Appendix G}
\renewcommand{\theequation}{G.\arabic{equation}}

In \cite{8007328}, the authors derived a BER floor for the ED under perfect TS. This has shown the existence of a BER floor in AmBC. In what follows, we use mathematical analysis to explain why the BERs in Fig. \ref{fig4} and Fig. \ref{fig6} suffer from BER floors under RTSE.

By substituting \eqref{21} into the first term of \eqref{13}, we obtain \eqref{G1}, as shown at the top of this page.

Assuming $\left| h \right| \le \left| \mu  \right|$ and substituting $\sigma _{\min }^2 = {\left| h \right|^2}{P_s} + {N_\omega }$ and $\sigma _{\max }^2 = {\left| \mu  \right|^2}{P_s} + {N_\omega }$ into \eqref{G1}, we obtain \eqref{G2}, as shown at the top of next page, where $SNR = \frac{{{P_s}}}{{{N_\omega }}}$.
\begin{figure*}[t]
\normalsize
\begin{align}\label{G2}\notag
\frac{{\gamma _{{\rm{th,nopt}}}^{{\rm{ip}}} - N\sigma _{\min }^2}}{{\sqrt N \sigma _{\max }^2}} &= \frac{1}{2}\frac{{\left( {N - {n_a}} \right)\left( {{{\left| \mu  \right|}^2} - {{\left| h \right|}^2}} \right)}}{{\left( {\sqrt N \left( {{{\left| h \right|}^2} + \frac{1}{{SNR}}} \right) + \sqrt {{n_a}{{\left( {{{\left| h \right|}^2} + \frac{1}{{SNR}}} \right)}^2} + \left( {N - {n_a}} \right){{\left( {{{\left| \mu  \right|}^2} + \frac{1}{{SNR}}} \right)}^2}} } \right)}}\\ \notag
&+ \frac{1}{2}\frac{{\left( {{{\left| \mu  \right|}^2} - {{\left| h \right|}^2}} \right)\left( {\sqrt N \sqrt {{n_a}{{\left( {{{\left| \mu  \right|}^2} + \frac{1}{{SNR}}} \right)}^2} + \left( {N - {n_a}} \right){{\left( {{{\left| h \right|}^2} + \frac{1}{{SNR}}} \right)}^2}}  - {n_a}\left( {{{\left| \mu  \right|}^2} + \frac{1}{{SNR}}} \right)} \right)}}{{\left( {{{\left| h \right|}^2} + \frac{1}{{SNR}}} \right)\left( {\sqrt {{n_a}{{\left( {{{\left| \mu  \right|}^2} + \frac{1}{{SNR}}} \right)}^2} + \left( {N - {n_a}} \right){{\left( {{{\left| h \right|}^2} + \frac{1}{{SNR}}} \right)}^2}}  + \sqrt N \left( {{{\left| \mu  \right|}^2} + \frac{1}{{SNR}}} \right)} \right)}},\tag{G.2}
\end{align}
\vspace{-15pt}
\hrulefill
\end{figure*}

As SNR approaches infinity, \eqref{G2} reaches an error floor at
\vspace{-15pt}
\begin{align}\label{G3}\notag
&\frac{{\gamma _{{\rm{th,nopt}}}^{{\rm{ip}}} \!\!-\!\! N\sigma _{\min }^2}}{{\sqrt N \sigma _{\max }^2}} \!=\! \frac{{\frac{1}{2}\left( {N - {n_a}} \right)\left( {{{\left| \mu  \right|}^2} - {{\left| h \right|}^2}} \right)}}{{\left( {\sqrt N {{\left| h \right|}^2} \!\!+\!\! \sqrt {{n_a}{{\left| h \right|}^4} \!\!+\!\! \left( {N \!-\! {n_a}} \right){{\left| \mu  \right|}^4}} } \right)}}\\ \notag
 &+ \frac{{\frac{1}{2}\left( {{{\left| \mu  \right|}^2} \!\!-\!\! {{\left| h \right|}^2}} \right)\left( {\sqrt N \sqrt {{n_a}{{\left| \mu  \right|}^4} \!\!+\!\! \left( {N \!\!- \!\!{n_a}} \right){{\left| h \right|}^4}} \!\! - \!\!{n_a}{{\left| \mu  \right|}^2}} \right)}}{{{{\left| h \right|}^2}\left( {\sqrt {{n_a}{{\left| \mu  \right|}^4} + \left( {N - {n_a}} \right){{\left| h \right|}^4}}  + \sqrt N {{\left| \mu  \right|}^2}} \right)}}.\tag{G.3}
\end{align}
\vspace{-10pt}

Since $Q( x )$ approaches zero if and only if $x \to \infty $, \eqref{G3} indicates that the first term of \eqref{13} is larger than zero, leading to an error floor. The same conclusion can be obtained when $\left| h \right| > \left| \mu  \right|$. This analysis can also be applied to other terms in \eqref{13}, resulting in the presence of BER floors in the overall BER performance. Consequently, both \eqref{13} and \eqref{14} exhibit BER floors, which are evident in Fig. \ref{fig4} and Fig. \ref{fig6}.

\color{black}

\ifCLASSOPTIONcaptionsoff
  \newpage
\fi
\bibliographystyle{IEEEtran}
\bibliography{refa}
\vspace{-20pt}

\end{document}